\documentclass[prd,aps,
longbibliography,
preprintnumbers,
twocolumn,
nofootinbib,
floatfix,
superscriptaddress,amsmath,amssymb]{revtex4-2}

\usepackage{graphicx}
\usepackage{xcolor}
\usepackage{dcolumn}
\usepackage{bm}
\usepackage[T1]{fontenc} 
\usepackage{epsfig}
\usepackage{rotating}
\usepackage{amssymb}
\usepackage{dsfont}
\usepackage{psfrag}
\usepackage{amsmath,euscript,array,mathrsfs}
\usepackage{bbold}
\usepackage{epsf}
\usepackage{float}
\floatstyle{plaintop}
\usepackage[utf8]{inputenc}
\usepackage{placeins}
\usepackage{orcidlink}
\usepackage{soul}
\usepackage[caption=false]{subfig}

\newcommand{\beq}{\begin{equation}}
\newcommand{\eeq}{\end{equation}}
\newcommand{\beqs}{\begin{eqnarray}}
\newcommand{\eeqs}{\end{eqnarray}}

\newcommand{\orcidauthorPIAI}{0000-0002-2251-0111}
\newcommand{\orcidauthorBENNETT}{0000-0002-1678-6701}
\newcommand{\orcidauthorLUCINI}{0000-0001-8974-8266}
\newcommand{\orcidauthorMASON}{0000-0002-1857-1085}
\newcommand{\orcidauthorRINALDI}{0000-0003-4134-809X}
\newcommand{\orcidauthorVADACCHINO}{0000-0002-5783-5602}

\begin{document}

\preprint{RIKEN-iTHEMS-Report-24}
\preprint{ET-0515A-24}

\title{The density of states method for symplectic gauge theories at finite temperature}

\author{Ed Bennett\,\orcidlink{\orcidauthorBENNETT}}
\email{e.j.bennett@swansea.ac.uk}
\affiliation{Swansea Academy of Advanced Computing, Swansea University, Fabian Way, SA1 8EN, Swansea, Wales, UK}
\author{Biagio Lucini\,\orcidlink{\orcidauthorLUCINI}}
\email{b.lucini@swansea.ac.uk}
\email{b.lucini@qmul.ac.uk}
\affiliation{Department of Mathematics, Faculty  of Science and Engineering,
Swansea University, Fabian Way, SA1 8EN Swansea, Wales, UK}
\affiliation{Swansea Academy of Advanced Computing, Swansea University,
Fabian Way, SA1 8EN, Swansea, Wales, UK}
\affiliation{School of Mathematical Sciences, Queen Mary University of London, Mile End Road, London, E1 4NS, UK}
\affiliation{Department of Mathematics, Faculty  of Science and Engineering,
Swansea University, Fabian Way, SA1 8EN Swansea, Wales, UK}
\author{David Mason\,\orcidlink{\orcidauthorMASON}}
\email{2036508@Swansea.ac.uk}
\email{d.mason@fz-juelich.de}
\affiliation{Department of Physics, Faculty  of Science and Engineering,
Swansea University,
Singleton Park, SA2 8PP, Swansea, Wales, UK}
\affiliation{Department of Mathematics, Faculty  of Science and Engineering,
Swansea University, Fabian Way, SA1 8EN Swansea, Wales, UK}
\affiliation{Jülich Supercomputing Centre, Forschungszentrum Jülich, D-52425 Jülich, Germany}
\author{Maurizio Piai\,\orcidlink{\orcidauthorPIAI}}
\email{m.piai@swansea.ac.uk}
\affiliation{Department of Physics, Faculty  of Science and Engineering,
Swansea University,
Singleton Park, SA2 8PP, Swansea, Wales, UK}
\author{Enrico Rinaldi\,\orcidlink{\orcidauthorRINALDI}}
\email{erinaldi.work@gmail.com}
\affiliation{Interdisciplinary Theoretical \& Mathematical Science Program, RIKEN (iTHEMS), 2-1 Hirosawa, Wako, Saitama, 351-0198, Japan}
\author{Davide Vadacchino\,\orcidlink{\orcidauthorVADACCHINO}}
\email{davide.vadacchino@plymouth.ac.uk}
\affiliation{Centre for Mathematical Sciences, University of Plymouth, Plymouth, PL4 8AA, United Kingdom}
 
\date{\today}%
\begin{abstract}

We study the finite-temperature behaviour of the $Sp(4)$ Yang-Mills lattice theory in four dimensions,
by applying the Logarithmic Linear Relaxation (LLR) algorithm.
We demonstrate %
the presence of coexisting (metastable) phases, when the system is in the proximity of  the transition.
We measure observables such as 
the free energy, 
the expectation value of the plaquette operator and of the Polyakov loop,
as well as the specific heat, and the Binder cumulant.
We use these results to obtain a high-precision measurement of the critical coupling at the 
confinement-deconfinement transition, and assess its systematic uncertainty,
for one value of the lattice extent in the time direction.
 Furthermore, we perform an extensive study of the finite-volume behaviour 
of the lattice system, 
by repeating the measurements for fixed lattice time extent, 
while increasing the spatial size of the lattice.
We hence characterise the  first-order transition on the lattice,
and present the first results in the literature on this theory for the infinite volume extrapolation
of lattice quantities related to latent heat and interface tension.

Gauge theories with $Sp(4)$ group have been proposed as
 new dark sectors
to provide a fundamental origin for the current phenomenological evidence of dark matter.
A phase transition at high temperature, in such a new dark sector,
occurring in the early universe, 
might have left a relic stochastic background of gravitational waves.
Our results represent a milestone toward 
establishing whether such a new physics signal is
 detectable in future experiments, as they enter the calculation of 
  the parameters, $\alpha$ and $\beta$,
controlling the power spectrum of gravitational waves.
We also outline the process needed in the
 continuum extrapolation of our measurements, and test its feasibility on one additional choice
 of temporal extent of the lattice.

\end{abstract}

\maketitle

\tableofcontents

\section{Introduction}

First-order phase transitions play a central role in our understanding of fundamental physics. 
They are particularly important in 
the context of the thermal history of the early universe, and its relation  to
 extensions of the Standard Model of particle physics.
Nucleation and expansion of bubbles that result from first-order transitions have striking 
implications. They  can
provide the out-of-equilibrium condition satisfying one of the three  requirements identified by Sakharov~\cite{Sakharov:1967dj} to explain, via baryogenesis, the matter-antimatter asymmetry in the observable universe.
Furthermore, bubble dynamics triggered by  first-order phase transitions may lead to a 
 relic stochastic background of gravitational waves~\cite{Witten:1984rs,Kamionkowski:1993fg,Allen:1996vm,Schwaller:2015tja,
 Croon:2018erz,Christensen:2018iqi}, which is testable in 
future experiments~\cite{Seto:2001qf,
 Kawamura:2006up,Crowder:2005nr,Corbin:2005ny,Harry:2006fi,
 Hild:2010id,Yagi:2011wg,Sathyaprakash:2012jk,Thrane:2013oya,
 Caprini:2015zlo,
 LISA:2017pwj,
 LIGOScientific:2016wof,Isoyama:2018rjb,Baker:2019nia,
 Brdar:2018num,Reitze:2019iox,Caprini:2019egz,
 Maggiore:2019uih}---see also Ref.~\cite{afzal2023nanograv}.

The thermal history of the Standard Model of particle physics proceeds via two smooth cross-overs. Lattice simulations indicate that  in quantum chromodynamics (QCD), with physical quark masses, deconfinement happens via a cross-over~\cite{Aoki:2006we}.  The electroweak transition, in which the $SU(2)_L\times U(1)_Y$ gauge symmetry of the Standard Model breaks  to the $U(1)$ of quantum electrodynamics (QED), has also been shown to be a cross-over, by a combination of perturbative  and
 non-perturbative results~\cite{Kajantie:1996mn,Karsch:1996yh,Gurtler:1997hr,
Rummukainen:1998as,Csikor:1998eu,Aoki:1999fi,
DOnofrio:2015gop}---see  Refs.~\cite{Laine:1998jb,Morrissey:2012db} for reviews, and also Ref.~\cite{Gould:2022ran}
for a recent non-perturbative update. This finding is particularly disappointing, as
it leads to the failure, within the Standard Model, of electroweak baryogenesis.

The past two decades saw the development of a whole branch of literature on new dark sectors---see for instance
Refs.~\cite{Strassler:2006im,Cheung:2007ut,Hambye:2008bq,Feng:2009mn,
Cohen:2010kn,Foot:2014uba,
Bertone:2016nfn}. New matter fields carrying new strongly coupled interactions are postulated to exist, 
with only feeble
couplings to the Standard Model, which allow such a dark sector and its particle content 
 to have escaped experimental detection.
In combination with the growing body of experimental evidence of dark
 matter (see the review in Ref.~\cite{Cirelli:2024ssz}, and references therein), this has led to the formulation of
 a plethora of  models realising this scenario.
Depending on the details of the mechanism implemented, they may be referred to
as composite dark matter models, such as those in
Refs.~\cite{DelNobile:2011je,
Hietanen:2013fya,Cline:2016nab,Kribs:2016cew,Cacciapaglia:2020kgq,Dondi:2019olm,
Ge:2019voa,Beylin:2019gtw,Yamanaka:2019aeq,Yamanaka:2019yek,Cai:2020njb},
or strongly interacting massive particle (SIMP) models, such as  those in 
Refs.~\cite{Hochberg:2014dra,Hochberg:2014kqa,Hochberg:2015vrg,Bernal:2017mqb,Berlin:2018tvf,
Bernal:2019uqr,Tsai:2020vpi,Kondo:2022lgg,Chu:2024rrv}---see also the
literature on dark
 $Sp(4)$ gauge theories~\cite{Maas:2021gbf,Zierler:2021cfa,Kulkarni:2022bvh}, or the recent proposals in Refs.~\cite{Pomper:2024otb,Appelquist:2024koa}, and references therein, or stealth dark matter models~\cite{Appelquist:2015yfa,Appelquist:2015zfa}, or dark baryon models~\cite{Francis:2018xjd}.

If any such new, strongly coupled, dark sectors play a role in the early universe, then it is 
of central importance to establish whether their finite-temperature 
dynamics leads to a  first-order phase transition,
and furthermore to achieve a precise characterisation of the physics in proximity of such a transition.
By doing so, one can, in principle,
determine those parameters that 
enter the calculation of
the power spectrum of the relic stochastic background of gravitational waves.
Borrowing  notation from Ref.~\cite{Caprini:2019egz}, the two prominent example are
$\alpha$ and $\beta/H_{\ast}$, that are related to  latent heat and  bubble nucleation rate, respectively.\footnote{The calculation of the  nucleation rate presents its own intrinsic challenges~\cite{Kalikmanov2013}.}
(They are used as input values in numerical packages such as
PTPlot~\cite{Caprini:2019egz}.)
The highly challenging goal of measuring these parameters in strongly coupled theories
is currently being pursued via several 
 complementary investigation strategies---see the discussions in
Ref.~\cite{Huang:2020crf}, and in
Refs.~\cite{Halverson:2020xpg,Kang:2021epo,Reichert:2021cvs,Reichert:2022naa,Pasechnik:2023hwv}.
 Lattice numerical results can be supplemented by effective tools as
 the Polyakov-loop~\cite{
 Pisarski:2000eq,
Pisarski:2001pe,Pisarski:2002ji,Sannino:2002wb,
Ratti:2005jh,Fukushima:2013rx,Fukushima:2017csk,Lo:2013hla,Hansen:2019lnf}
or 
matrix 
models~\cite{
Meisinger:2001cq,Dumitru:2010mj,
Dumitru:2012fw, Kondo:2015noa,Pisarski:2016ixt,Nishimura:2017crr,Guo:2018scp,
KorthalsAltes:2020ryu,Hidaka:2020vna}.

The finite-temperature behaviour of lattice gauge theories 
is the subject of intense scrutiny.
For a review of the current status of $SU(3)$ gauge theories, see the discussion in Ref.~\cite{borsanyi:2022xml} and  the review~\cite{Aarts:2023vsf} (as well as
Refs.~\cite{Svetitsky:1982gs,Yaffe:1982qf}), Refs.~\cite{Kajantie:1981wh,Celik:1983wz,Kogut:1983mn,
Svetitsky:1983bq,Gottlieb:1985ug,Brown:1988qe,Fukugita:1989yb,Bacilieri:1989ir,
Alves:1990pn,
Boyd:1995zg,Boyd:1996bx,
Borsanyi:2012ve,Shirogane:2016zbf} for pure gauge, and 
Refs.~\cite{Saito:2011fs,Ejiri:2019csa,Kiyohara:2021smr,
Fromm:2011qi,Cuteri:2020yke,
Borsanyi:2021yoz} for heavy quarks; for theories with other gauge groups,
see for instance Refs.~\cite{Lucini:2002ku,Lucini:2003zr,Lucini:2005vg,Panero:2009tv,Datta:2010sq,Lucini:2012wq}
 for $SU(N)$, Ref.~\cite{Pepe:2005sz,Pepe:2006er,Cossu:2007dk,Bruno:2014rxa} for $G_2$,
Ref.~\cite{Appelquist:2015yfa,Appelquist:2015zfa,LatticeStrongDynamics:2020jwi} for stealth dark matter with $SU(4)$ gauge group, and Ref.~\cite{Holland:2003kg}
 for pioneering work on $Sp(2N)$ gauge theories---see also the recent work Ref.~\cite{Bruno:2024dha}. 

It is worth mentioning that an alternative route to the treatment of strongly coupled
gauge theories has become available with the establishment of  gauge-gravity dualities~\cite{Maldacena:1997re,Gubser:1998bc,Witten:1998qj,Aharony:1999ti}.
In special cases, the gravity description can capture  the salient features of
confinement and chiral symmetry breaking in the dual field theories~\cite{Witten:1998zw,
 Klebanov:2000hb,Maldacena:2000yy,Chamseddine:1997nm,Babington:2003vm,Butti:2004pk,
 Brower:2000rp,Karch:2002sh,Kruczenski:2003be,Sakai:2004cn,Sakai:2005yt}. The holographic approach to the treatment of strong dynamics has been applied to first-order phase transitions and the associated 
 out-of-equilibrium dynamics, for example in Refs.~\cite{Bigazzi:2020phm,Ares:2020lbt,Bea:2021zsu,Bigazzi:2021ucw,Henriksson:2021zei,Ares:2021ntv,Ares:2021nap,Morgante:2022zvc}.
We do not further discuss this  approach, but refer the reader to the 
literature on the subject for details and insight.

Unfortunately, the emergence of first-order phase transitions presents a unique challenge to conventional lattice simulations. Metastable dynamics near criticality results in Monte Carlo (Markov chain) update methods   
getting stuck in one of possible multiple vacua, during
the ensemble generation process. Specifically, if we call $\tau$ the Monte Carlo time needed by the system to tunnel between two vacua, while $N_{\mathrm{conf}}$ is the number of generated configurations,  the condition $\tau \ll N_{\mathrm{conf}}$  must be fulfilled 
in order to obtain the correct sampling of the equilibrium distribution.
 This requirement leads to a dramatic increase in the number of generated configurations as the transition becomes stronger.  In the approach to the infinite volume limit, in which the potential barrier separating phases grows with the system size, this critical slowing down may become an intractable numerical problem~\cite{Berg:1991cf}. 

Inspired by flat histogram~\cite{Berg:1991cf} and density of states~\cite{Wang:2000fzi} methods, the Logarithmic Linear Relaxation (LLR) algorithm~\cite{Langfeld:2012ah,Langfeld:2013xbf,Langfeld:2015fua,Cossu:2021bgn} offers a
new opportunity to perform numerical studies that can be improved systematically and controllably, 
even in the presence of metastability, phase coexistence, tunnelling, and other phenomena
naturally associated with first-order phase transitions.
We recall the physics principles and algorithmic processes underlying the LLR in the body of the paper.
The application of the LLR algorithm to Abelian
 gauge theories has been studied in detail~\cite{Langfeld:2015fua}.
In the non-Abelian case, the properties of $SU(3)$ have been investigated  at
 zero-temperature~\cite{Cossu:2021bgn}, while at finite-temperature 
 some of the authors of this work have produced the first physical results, published in Refs.~\cite{Mason:2022trc,Mason:2022aka,Lucini:2023irm}.
Finite-temperature studies exist also for $SU(4)$~\cite{Springer:2021liy,
   Springer:2022qos} and general $SU(N)$~\cite{Springer:2023wok,Springer:2023hcc}.  

In this paper, we consider the $Sp(4)$ pure gauge theory.
Following up
from the minimal composite Higgs model proposal in Ref.~\cite{Barnard:2013zea}---see also the tables in Refs.~\cite{Ferretti:2013kya,Ferretti:2016upr,Cacciapaglia:2019bqz}---and from the minimal SIMP dark-matter model proposal in
Ref.~\cite{Hochberg:2014kqa}, 
interest in new gauge theories with $Sp(2N)$ gauge group (in particular for $Sp(2)\sim SU(2)\sim SO(3)$ and $Sp(4)\sim SO(5)$) has motivated the development of the research programme of Theoretical Explorations on the Lattice 
with Orthogonal and Symplectic groups (TELOS)~\cite{Bennett:2017kga,Bennett:2019jzz,Bennett:2019cxd,Bennett:2020hqd,Bennett:2020qtj,
Bennett:2022yfa,Bennett:2022gdz,Bennett:2022ftz,
Bennett:2023wjw,
Bennett:2023gbe,Bennett:2023mhh,
Bennett:2023qwx,Bennett:2024cqv,Bennett:2024wda}---see also Refs.~\cite{Lee:2018ztv,Lucini:2021xke,Bennett:2021mbw,Hsiao:2022gju,Hsiao:2022kxf}, as well as  Refs.~\cite{Maas:2021gbf,Zierler:2021cfa,
Kulkarni:2022bvh,Zierler:2022qfq,Zierler:2022uez,Bennett:2023rsl,Dengler:2023szi,Dengler:2024maq} 
for dark matter models  based on $Sp(4)$ gauge theories.
It is hence natural to seek to characterise these theories at high temperature, starting from the Yang-Mills theory~\cite{Mason:2023ixv}, to go beyond the seminal lattice work in Ref.~\cite{Holland:2003kg}.

In particular, in the following we provide the first direct lattice calculation of the non-perturbative features of the effective potential and free energy of the $Sp(4)$ theory near its confinement-deconfinement phase transition. This is input information for future  calculations of out-of-equilibrium dynamical quantities, such as the nucleation rate and the relic stochastic gravitational-wave power spectrum, as anticipated.
To this purpose, we develop the necessary formalism, highlight the essential  features of the LLR mechanism that make this measurement possible, and perform an extensive   finite volume study.

The paper is organised as follows.
We  define our theory in Sect.~\ref{Sec:Lattice}, with particular reference to the lattice variables and action.
Section~\ref{sec:LLR} provides a self-contained, brief description of the LLR algorithm, as applied to the 
theory of interest.
The main body of the paper starts with Sect.~\ref{Sec:Thermo}, in which we explain how to characterise the thermodynamics of the system in proximity of the phase transition, and how we can extract the 
physical quantities of interest.
Our numerical results are presented and critically discussed in Sect.~\ref{sec:Results}.
A short outlook in Sect.~\ref{Sec:outlook} closes the paper, except that we defer a number of important and useful technical details to the Appendix.


\section{Lattice setup}
\label{Sec:Lattice}
The continuum, Yang-Mills theory of interest has $Sp(4)$ gauge group, and lives  in four space-time dimensions. 
The field content consists only of the 
gauge fields, with non-Abelian  self-couplings dictated by the minimal gauge principle.
To study it on the lattice,  we, first, Wick rotate the Minkowski space-time and, second, discretise the resulting Euclidean space-time onto a hyper-cubic lattice, with spacing denoted as $a$, and (hyper-)volume $\tilde{V} \equiv a^4 (N_t \times N_s^3)$, where $N_{t(s)}$ stands for the number of sites in  the time (space) direction. 
We label the sites of the lattice by $(n_t, \vec{n}_s)$, where $n_t=0,\,\cdots,\,N_t-1$ and $\vec{n}_s= \{ n_i  | i = 1, 2, 3 \}$, with $n_i = 0,\,\cdots,\,N_s -1$. 

The lattice link variables, $U_\mu (n_t, \vec{n}_s)\in Sp(4) \subset SU(4)$, where $\mu$ denotes the space-time direction, are matrix valued in the  $Sp(4)$ group, hence they obey the relation $U_\mu \Omega (U_\mu)^T = \Omega$, with the symplectic matrix, $\Omega$,  given by 
\beq
\Omega \equiv \begin{pmatrix}
0 & \mathds{1}_{2\times 2}\\
-\mathds{1}_{2\times 2} & 0
\end{pmatrix}\,.
\eeq 
This relation restricts the elements of the group
to obey a block-matrix  relation  discussed in Appendix~\ref{App:HB_Append}.

Observables are defined as vacuum expectation values (VEVs) of operators, $O=O[U] $, 
that depend on the configuration of all the link variables (generically denoted as $U$). They are
 calculated as weighted averaged defined by the  path integral:
\beq
\label{eqn:Observable}
\langle O \rangle_\beta \equiv  \frac{1}{Z(\beta)} \int [DU_\mu] O[U] e^{-\beta S[U]}\,,
\eeq
where the integral is over all possible configurations, $D U_\mu$ is the Haar measure,
$S[U]$ is the classical action evaluated on a configuration,
 and $\beta\equiv 8/g_0^2$ is the bare coupling. 
The partition function is defined as
\beq
\label{eqn:Partition}
Z(\beta) \equiv  \int [DU_\mu]e^{-\beta S[U]}\,.
\eeq
For this study, we elect to adopt the  standard (unimproved) Wilson action:
\beq
\label{eqn:WilsonAction}
S[U]\equiv\frac{6\tilde{V}}{a^4}(1-u_p[U])\,,
\eeq
where  $u_p[U]$ is the average plaquette evaluated on one lattice configuration, $U$.

We consider a lattice with $N_t \ll N_s$, and impose periodic boundary conditions in the temporal direction. Such a set up can be interpreted in terms of a thermalised, equilibrium system at temperature, $T$, determined by 
 the inverse of the temporal extent, $T\equiv 1/(a N_t)$. For a fixed lattice size,  the temperature can be altered 
 by varying the lattice spacing, $a(\beta)$,  via its dependence on the lattice coupling, $\beta$. If a phase transition takes place in the system considered at fixed temporal extension, $N_t$, its critical coupling, $\beta_C$,  is  related to the critical temperature as $T_C \equiv 1/(a(\beta_C) N_t)$.  While these two quantities, $T_C$ and $\beta_C$, can, in principle, be traded for one another, by using the aforementioned relation, in practice it is convenient to study the transition in terms of $\beta_C$, at fixed $N_t$. One then recovers the physical quantity, $T_C$, by taking the continuum limit in a way that implements also a scale-setting procedure. This can be done by building dimensionless ratios of physical observables, for example  $T_c/\sqrt{\sigma}$, with $\sigma$ the string tension.    

The deconfinement phase transition is associated with the spontaneous breaking of the centre symmetry of the group,
which for all $Sp(2N)$ groups is $\mathbb{Z}_2$. The order parameter associated to this transition is the Polyakov loop, a closed loop of fundamental link variables wrapping the periodic temporal direction,  normalised as follows:
\beq
\label{eqn:PolyakovLoop}
l_p[U]  \equiv \frac{1}{4 N_s^3} \sum_{\vec{n}_s} \textrm{Tr}\left(\prod_{n_t = 0}^{N_t - 1} U_0(n_t, \vec{n}_s)\right)\,. 
\eeq
In the confined phase, this observable vanishes, while in the deconfined phase it assumes a finite value, $l_p[U] = z |l_p[U]| \neq 0$, where $z\in \mathbb{Z}_2$ is an element of the centre of the group, therefore $z=\pm 1$. 

In 
commonly used 
algorithms for lattice field theory calculations, 
vacuum expectation values are computed by replacing the path integral with an ensemble average,  the configurations composing the ensemble having been generated by using importance sampling methods. In the proximity of first-order phase transitions, such algorithms are affected by metastable dynamics, so that the Monte Carlo correlation time, $\tau$,  is expected to scale exponentially with the size of the system. For a system in $d$ dimensions, and at leading order in the linear size, $L$, one expects $\tau$ to scale as
 follows (see, e.g., the discussion in Ref.~\cite{Berg:1992qua}):
\begin{eqnarray}
\label{eq:tunnelingtime}
\tau \propto L^{\alpha} e^{f L^{d- 1}} \,.
\end{eqnarray}
In Eq.~(\ref{eq:tunnelingtime}), $\alpha$ and $f$ (the latter is related to the tension of the interface between the two vacua) are positive constants, dictated by the dynamics. This exponential increase of $\tau$ leads to uncontrolled systematic errors, and the resource-ineffective need to produce unmanageably large ensembles. In order to alleviate this difficulty, we utilise the LLR method, defined in the next section.

\section{Density of states method and LLR algorithm}\label{sec:LLR}
The aim of the LLR algorithm is to estimate efficiently and accurately  the density of states, 
$\rho(E)$, defined as
\beq
\label{eqn:DoS}
\rho(E)\equiv\int [ D U_\mu] \delta(S[U]-E)\,,
\eeq
by sampling the space of configurations of the system,  $U$,  as a function of the energy, $E$.
This quantity counts the number of states with a given energy (action), allowing 
to recast the calculation of the partition function 
as a one-dimensional integral over the energy,
\beq
\label{eqn:ActionDS}
Z(\beta)=\int d E \rho(E) e^{-\beta E}\,. 
\eeq
A similar reformulation applies to ensemble averages of observables that depend  explicitly
only on the energy:
\beq
\label{eq:vev_obs}
 \langle O \rangle_\beta =\frac{1}{Z(\beta)}\int dE \rho(E) O(E) e^{-\beta E} \,. 
\eeq
The average plaquette, $u_p$, is one such observable, to which we return shortly.

The numerical extraction of the density of states, ${\rho}(E)$, in the LLR method,
starts by approximating it as a piecewise log-linear function, $\tilde{\rho}(E)$. 
The energy range of interest
 is divided into energy intervals, $E_n -\Delta_E/4 < E < E_n + \Delta_E/4$, with,
\beq
\label{eq:piecewise}
\log \tilde{\rho}(E) \equiv a_n \left(E - E_n \right) + c_n\,,
\eeq
for $n=1,\, \cdots,\, N_{\rm int}$. For sufficiently small choices of $\Delta_E$, up to an exponentially suppressed error, the coefficients $a_n$ and $c_n$ must be such that  $\tilde{\rho}(E) \simeq \rho(E)$.  In the limit $\Delta_E \to 0$, one recovers the exact density of  states, $\rho(E)$. The quantities $c_n$ denote the value of $\log \tilde{\rho}(E)$ at the centre of the intervals. By continuity of $\tilde{\rho}(E)$, they satisfy the relations
\beq
\label{eq:c_n}
c_n =  c_1 + \frac{\Delta_E}{4} a_1 + \frac{\Delta_E}{2} \sum_{k=2}^{n-1} a_k +
\frac{\Delta_E}{4} a_n \,.
\eeq
Furthermore, we set $c_1 = 0$, as in most cases this factor will not affect the results. (We treat 
separately the observables for which this statement does not apply, such as the free energy.) 
In order to determine the coefficients $a_n$, we follow our previous work~\cite{Lucini:2023irm}. 
The  process we employ requires the computation of the ensemble average of a known,
appropriately chosen observable,
and then to dial the coefficients, $a_n$, in order to reproduce the same result, 
with the  replacement of $\tilde{\rho}$ in place of $\rho$, in Eqs.~(\ref{eqn:ActionDS}) and~(\ref{eq:vev_obs}).
The numerical algorithm is described in detail in Appendix~\ref{App:LLR} and~\ref{App:HB_Append}. We repeat our determination of the coefficients on different stochastic samples and bootstrap the results to estimate uncertainties. To control the systematic error due to the finiteness of the interval, $\Delta_E$, we repeat our calculations for multiple choices of  $\Delta_E$ sizes, and extrapolate to the $\Delta_E \to 0$ limit.

The partition function is approximated, as a function of the coefficients, $a_n$ and $c_n$,
by the expression
\begin{eqnarray}
\label{eqn:ActionDoS}
Z(\beta) &\simeq& 
\sum_{n=1}^{N_{\rm int}} e^{-a_n E_n + c_n} \int_{E_n-\Delta_E/4}^{E_n+\Delta_E/4}d E e^{(a_n -\beta) E} 
\,. 
\end{eqnarray}
We can estimate expectation values of operators $O$  as
\beq
\label{eqn:ObsDoS}
\langle O \rangle_\beta \simeq \sum_{n=1}^{N_{\rm int}} \frac{e^{-a_n E_n + c_n}}{Z(\beta)}  \int_{E_n-\Delta_E/4}^{E_n+\Delta_E/4}d E O(E) e^{(a_n -\beta) E}\,.
\eeq
The probability distribution for the energy, or, equivalently, for the average plaquette, $u_p$, is given by
\beqs
\label{eqn:PlaqDistribution}
P_\beta (u_p ) &\equiv& \frac{1}{Z(\beta)}\rho(E)e^{-\beta E}
\Big|_{E=\frac{6\tilde{V}}{a^4}(1-u_p)}\,,
\eeqs
and again is approximated by replacing $\rho$ with $\tilde{\rho}$.

Once the values of  $a_n$ are known, one can  sample the space of configurations with general weight,
not necessarily the Boltzmann weight appearing in the canonical ensemble.
In particular, we can generate ensemble averages with  flat energy distribution within  narrow energy  intervals---closely related to the microcanonical ensemble, but for subtleties that we discuss in the Appendix. By measuring an observable, $B[U]$, (for example, the Polyakov loop) on these configurations, we can compute the expectation values, through the equation
\beq
\label{eqn:vev_gen}
 \langle B[U] \rangle_\beta = \frac{1}{Z(\beta)} \sum_{n=1}^{N_{\rm int}}
 \frac{\Delta_E}{2} \tilde{\rho}(E_n) \tilde{B}[U] \ , 
\eeq
 where $\tilde{B}[U]$ is the vacuum expectation value for configurations restricted to the energy interval of the observable, multiplied by a weight factor,
\beq
\label{eqn:vev_gen_small}
\tilde{B}[U] \equiv \left\langle \left\langle B[U]e^{-\beta S[U] + a_n
     (S[U] - E_n)} \right\rangle \right\rangle _n(a_n) \,.
\eeq

\section{Thermodynamics}
\label{Sec:Thermo}
In this Section, we provide an overview of the thermodynamic observables studied in our investigation. We discuss the overall strategy we adopt to characterise the phase transition of the $Sp(4)$ pure gauge theory. 
General expectations suggest that the order of the phase transition should change from second to first when the number of degrees of freedom of the system exceeds a certain threshold. As the $SU(3)$ gauge theory, which has only marginally fewer degrees of freedom than $Sp(4)$, has a well-established first-order deconfinement phase transition, it is natural to expect the $Sp(4)$ 
Yang-Mills theory to exhibit a first-order phase transition as well, as indeed confirmed by
pioneering investigations~\cite{Holland:2003kg}. In our discussion we concentrate on the characterisation of first-order phase transitions, with the expectation (to be checked \emph{a posteriori}) that such characterisation be supported by  our numerical results.

A signature of first-order phase transitions is the presence of a latent heat: a difference in internal energy between inequivalent equilibrium states coexisting at the critical temperature, $T_C$. It can be extracted from the discontinuity of the internal energy density, $\varepsilon(T)$, of the system, which is given by 
\beq 
\label{eq:internalE1}
\varepsilon(T) \equiv \frac{\kappa T^2}{V} \left (  \frac{\partial \ln Z(T)}{\partial T}\right) _V\,, 
\eeq
evaluated at the transition temperature, $T\rightarrow T_C$. 
In this expression, the partial derivative and the subsequent limit are computed at fixed volume.
Hence, direct calculation of the energy density on the lattice would require the use of  anisotropic lattices, to compute quantities known as Karsch coefficients~\cite{Karsch:1982ve}. Yet, if we assume the  pressure to be continuous across the transition, following Ref.~\cite{Lucini:2005vg}, the latent heat can be related to the plaquette discontinuity at the critical point, 
\beq
\label{eq:latheat}
\frac{L_h}{T_c^4} = - \left( 6 N_t^4 a
\frac{\partial\beta}{\partial a} 
{\Delta \langle u_p \rangle_\beta }\right)_{\beta=\beta_c}
{}\,.
\eeq
We are going to follow this approach, which is consistent with the large-$N$ arguments in Ref.~\cite{Panero:2009tv}.

In view of the potential presence of systematic effects due to how the manifestation of metastability affect the numerical study of first-order phase transitions, it is useful to consider multiple independent determinations of the same quantities, as well as the verification of known constraints among physical observables. Both strategies exploit the redundancy of the numerical measurements as a way to either cross-validate the results or provide an estimate of systematic effects. For instance, the specific heat, $C_V$, and the latent heat, $L_h$, are linked in the thermodynamic limit.

 At fixed temperature and outside the critical region, we can model the plaquette fluctuations with a Gaussian probability distribution, 
\beq
\label{eq:gaussianapprox}
P_\beta(u_p) = A \sqrt{\frac{(\tilde V / a^4)}{C}} \exp\left( -\frac{(u_p - u_{p,\,0})^2 \beta^2 (\tilde V / a^4) )}{2 C}\right) \,,
\eeq
where $C$ is the (infinite-volume) specific heat, $A$ is a normalisation constant, and $u_{p,\,0}$ is the average plaquette. It is understood that $C$ and $u_{p,\,0}$ are temperature-dependent. In the infinite volume limit, the distribution in Eq.~(\ref{eq:gaussianapprox}) becomes a Dirac delta function. In the transition region, ignoring mixed phase configurations,\footnote{These are defined by the existence of  macroscopic portions of the system realising  different phases, separated by interfaces. We  shall come back later to this more general case.} we can approximate the plaquette distribution as the sum of two Gaussians, corresponding to the two phases, 
\beq
\label{eq:doublegaussianapprox}
P_\beta(u_p) = P_\beta^{(+)}(u_p) + P_\beta^{(-)}(u_p) \,, 
\eeq
where $P_\beta^{(+)}(u_p)$ ($P_\beta^{(-)}(u_p)$) is described by parameters, $A^{(+)}$, $C^{(+)}$, $u_{p,\,0}^{(+)}$, ($A^{(-)}$, $C^{(-)}$, $u_{p,\,0}^{(-)}$), that generalise those appearing in Eq.~(\ref{eq:gaussianapprox}). 

From the linearity of the approximation in Eq.~(\ref{eq:doublegaussianapprox}), we infer that expectation values  of even powers of the plaquette are the sum of their expectation values in each of the two phases, 
\beq
\langle  u_p^{2k} \rangle =  \langle  u_p^{2k} \rangle_+ + \langle  u_p^{2k} \rangle_- \,, 
\eeq
where $\langle \dots \rangle_+$ denotes the expectation on the Gaussian parameterised by $A^{(+)}$, $C^{(+)}$, $u_{p,\,0}^{(+)}$ and $\langle \dots \rangle_-$ is the expectation on the Gaussian parameterised by $A^{(-)}$, $C^{(-)}$, $u_{p,\,0}^{(-)}$. This approximation becomes exact in the infinite volume limit, in which mixed phases are known to be absent. Therefore, for the specific heat,
\beq
\label{eq:specheat}
C_V \equiv \frac{6 \tilde V}{a^4} \left(\langle u_p^2 \rangle_\beta -\frac{}{} \langle u_p \rangle_\beta^2 \right) \,,
\eeq
we arrive at the expectation that
\beq
\label{eq:spec_to_latheat}
C_V^\mathrm{(max)} \xrightarrow[]{\frac{N_t}{N_s} \to 0} \frac{6 \tilde V}{4} \left( \Delta \langle u_p \rangle_{\beta_c}  \right)^2\,. 
\eeq

When the three-dimensional, spatial volume,  $V$, is finite, then $C_V$ has a maximum at a value of the coupling, that we denote as $\beta_{CV}(C_V)$.
 The value of the maximum diverges, in the approach to the thermodynamic limit,
  proportionally to the volume. 
This provides an operational way to determine the critical coupling, $\beta_C$, at fixed $N_t$:
\beq
\beta_C(N_t) = \lim_{N_s \to \infty} \beta_{CV}(C_V) \,.
\eeq

Any susceptibility (or susceptibility-like observable coupling to the degrees of freedom that are relevant for the transition) is expected to be characterised by a finite-volume peak or dip,
the value of which diverges in the thermodynamic limit. The position of the peak provides a finite-volume definition of a pseudo-critical coupling, that converges to the critical value in the infinite-volume limit. An additional energy-based observable  characterising phase transitions is the Binder cumulant~\cite{Challa:1986sk,Bhanot:1989wj}, defined as 
\beq
\label{eq:binder}
B_V(\beta) \equiv 1 - \frac{\langle u_p^4 \rangle_\beta}{3 \langle u_p^2 \rangle_\beta^2} \,. 
\eeq
For first-order phase transitions, in the thermodynamic limit,  a 
minimum appears at the critical coupling, $\beta_{CV}(B_V)$.\footnote{In the case of first order phase transitions, the value of the minimum in $B_V$ 
converges to a finite value in the infinite volume limit. By contrast, for continuous phase transitions
$B_V$ approaches a value of 2/3, and the 
 the dip around the critical coupling disappears. }

Similar arguments hold for the susceptibility of the Polyakov loop,
\beq
\chi_l (\beta) \equiv N_s^3 (\langle | l_p^2 | \rangle_\beta - \langle | l_p | \rangle_\beta^2) \,,
\eeq
which has a finite-volume maximum at the pseudo-critical value of the coupling, $\beta_{CV}(\chi_l)$. While at fixed $V$ the three pseudo-critical couplings we introduced, $\beta_{CV}(C_V)$, $\beta_{CV}(B_V)$, and $\beta_{CV}(\chi_l)$, are in general different, they must all converge to the same $\beta_C$ in the thermodynamic limit, at fixed $N_t$.

Another important observable characterising first-order phase transitions is the tension of the interface separating coexisting equilibrium phases. It can be extracted from the measurement  of the plaquette distribution. Specifically, at finite volume, if we consider the value of  $\beta$  for which the two peaks of the distribution have the same height, $P_{\mathrm{max}}$, the confinement-deconfinement interface tension, $\sigma_{cd}$, can be extracted from the relation
\beq
\label{eq:interface_probratio}
\frac{P_\mathrm{min}}{P_\mathrm{max}} \propto N_s^{\frac{1}{2}} \exp\left(-2 \frac{\sigma_{cd}}{T_C^3} \frac{N_s^2}{N_t^2} \right) \,,
\eeq
where $P_\mathrm{min}$ is the value of the probability at the minimum located between the maxima.
For a heuristic derivation of this expression, see, e.g., Ref.~\cite{Lucini:2005vg}. 

It is important to highlight that the contribution to the probability distribution of these interfaces is not captured by the double Gaussian approximation, Eq.~(\ref{eq:doublegaussianapprox}), as the former is due to a macroscopic, emergent interaction between subsystems, while the latter approximates the mixed-phase system in terms of two independent, free subsystems.
Nevertheless, the argument leading to Eq.~(\ref{eq:spec_to_latheat}) remains asymptotically correct. Note also that, in the derivation of Eq.~(\ref{eq:interface_probratio}) in Ref.~\cite{Lucini:2005vg}, effects such as the vibration of the interface are not considered; yet, these and other  effects we ignore, including the unknown proportionality factor, are expected to be suppressed as $N_s^{-2}$. Therefore, we can extract the interface tension from the quantity 
\beq
\tilde I \equiv -\frac{N_t^2}{2 N_s^2} \log \left ( \frac{P_\mathrm{min}}{P_\mathrm{max}}\right ) - \frac{N_t^2 }{4 N_s^2}\log(N_s) \,,
\label{eq:Interface0}
\eeq
by taking the limit $N_t/N_s \to 0$:
\beq
\label{eq:Interface}
\lim_{\frac{N_t}{N_s} \to 0} \tilde I  = \tilde \sigma_{cd} \,, 
\eeq
where, for convenience,  we have defined $\tilde \sigma_{cd} = \sigma_{cd} / T_C^3$.

The process of determining $\tilde{I}$ through the plaquette probability distribution in the critical region introduces a fourth definition of the finite-volume pseudo-critical $\beta$ as the value at which the peaks of the plaquette distribution are equal. We call this value $\beta_{CV}$.

\begin{table*}[th]
\caption{List of $Sp(4)$ lattice ensembles with extension in the time direction  $N_t=4$. 
For each of them, we list the lattice extension, $N_t \times N_s^3$,
 the number of (overlapping) subintervals, $N_{\rm int}$, 
the number of replicas, $n_R$, the extent  of the interval in average plaquette, $\Delta_{u_p}$,  and the maximum and minimum values of the plaquette, $(u_p)_{\mathrm{min}}$ and $(u_p)_{\mathrm{max}}$.
We also report the number of iterations defined in Appendix~\ref{App:LLR}: $\bar{m}$, 
$\tilde{m}$, $\hat{m}$, $n_{\text{Th}}$, $n_{M}$, and $n_{\text{fxa}}$. }
 \label{tab:lat_surf1}
\begin{center}
\begin{tabular}{|c|c|c|c|c|c|c|c|c|c|c|c|}
\hline
$N_t \times N_s^3$ & $N_{\mathrm{int}}$ & $n_R$ & $\Delta_{u_p}$ & $(u_p)_{\mathrm{min}}$ & $(u_p)_{\mathrm{max}}$ & $\bar{m}$ & $\tilde{m}$ & $\hat{m}$ & $n_{\mathrm{Th}}$ & $n_M$ & $n_{\mathrm{fxa}}$ \\
\hline
$4 \times 20^3$ & 48 & 20 & 0.00064 & 0.565 & 0.58 & 10 & 300 & 100 & 300 & 700 & 100 \\
$4 \times 20^3$ & 64 & 20 & 0.00048 & 0.565 & 0.58 & 10 & 300 & 100 & 300 & 700 & 100 \\
$4 \times 24^3$ & 48 & 20 & 0.00064 & 0.565 & 0.58 & 10 & 300 & 100 & 300 & 700 & 100 \\
$4 \times 24^3$ & 64 & 20 & 0.00048 & 0.565 & 0.58 & 10 & 300 & 100 & 300 & 700 & 100 \\
$4 \times 28^3$ & 64 & 20 & 0.00048 & 0.565 & 0.58 & 10 & 200 & 100 & 300 & 700 & 100 \\
$4 \times 28^3$ & 64 & 20 & 0.00025 & 0.568 & 0.576 & 10 & 200 & 100 & 300 & 700 & 100 \\
$4 \times 40^3$ & 96 & 25 & 0.00017 & 0.568 & 0.576 & 10 & 100 & 50 & 300 & 700 & 100 \\
$4 \times 40^3$ & 128 & 25 & 0.00013 & 0.568 & 0.576 & 10 & 100 & 50 & 300 & 700 & 100 \\
$4 \times 48^3$ & 128 & 25 & 0.00013 & 0.568 & 0.576 & 10 & 100 & 50 & 300 & 700 & 100 \\
\hline
\end{tabular}

\end{center}
\end{table*}

The LLR approach gives us access to thermodynamic observables that are exponentially (in the volume) hard to determine using importance sampling methods. Using analogies with statistical physics, we define the free energy, $F$, entropy, $s$, and temperature, $t$, of the microstates as
\beq
\label{eq:Therm_pot}
F\equiv E-ts,\,\,\,\, s = \log(\rho),\,\,\,\, t = \frac{\partial E}{\partial s} = \frac{1}{a_n},  
\eeq
where the internal energy, $E$, is given by the total action. As in classical thermodynamics, both $s$ and $E$ are state functions, defined up to an additive constant. In Eq.~(\ref{eq:c_n}), we have arbitrarily set the value of $c_1$, therefore the calculated value of the free energy will differ from the true value by a term linear in temperature. As we are mostly  interested in calculating the difference in the free energy at fixed temperatures, we can arbitrarily choose this value. We  define a reduced free energy, $f(t)$, to remove this first arbitrariness:  
\beq
\label{eq:Red_FE}
f(t) \equiv \frac{a^4}{\tilde{V}}(F(t)+\Sigma t)\,,
\eeq
with $\Sigma$  the average entropy of the microstates.  

As we shall see in our numerical results, with this definition the swallow-tail structure of $f(t)$, characteristic of first-order transitions, becomes  prominent. This structure is the result of physically different microstates coexisting at the same temperature, over a finite range of $t$, close to criticality. In addition to the unstable branch, the coexistence phenomenon gives rise to two metastable branches, corresponding, respectively, to the confined and deconfined states at temperatures where they have a higher free energy than the true equilibrium state.
We call $f_c^+$ the value of the reduced free energy evaluated at the point at which the two meta-stable branches cross. The value of $\beta$ at which this happens (at finite volume) gives rise to a fifth definition of the pseudo-critical temperature, which we call $\beta_{CV}(f)$. For further discussion of the structure of the free energy and our approach to determining it, see also our previous work, in Ref.~\cite{Lucini:2023irm}. By comparing the definition of Eq.~(\ref{eqn:PlaqDistribution}) with the free energy, we arrive at the relation
\beq
\exp\left(-\frac{F(t)}{t}\right) = Z(\beta) P_\beta(E)\bigg|_{\beta=1/t, E=F(t)+ts} \,,
\eeq
where $P_\beta(E)$ corresponds to the probability of being in a given micro-state with internal energy $E$, for $t=1/\beta$. Therefore, this definition of the critical coupling $\beta_{CV}(f)$, should be equivalent to $\beta_{CV}$ even at finite volume.


\section{Results}
\label{sec:Results}

This section contains the main body of the paper. We restrict attention to the characterisation of  the finite-temperature, deconfinement phase transition of the $Sp(4)$ Yang-Mills lattice theory. We consider a fixed extent of the temporal direction, $N_t = 4$, and study the extrapolation of our results towards the thermodynamic limit, by varying the number of sites in the spatial directions, $N_s=20,\,24,\,28,\,40,\,48$.

As a preliminary step, before carrying out a full, high precision, LLR analysis, we use standard importance sampling methods to determine the region of interest in the lattice coupling, $\beta$. By doing so, we set the plaquette range $(u_p)_\mathrm{min} < u_p < (u_p)_\mathrm{max}$ that will be explored by the LLR algorithm, and at the same time set initial values for the coefficients, $\{a_n^{(0)} \}_{n=1}^{N_{\rm int}}$, that will then be determined to high precision by our numerical, iterative processes. We carry out a standard importance sampling parameter scan of $\beta$ values around the critical coupling, for a lattice with extent $N_t\times N_s^3=4\times 20^3$. In Ref.~\cite{Holland:2003kg}, the critical inverse coupling for $N_t = 4$, in the thermodynamic limit, was determined to be $\beta_C=7.339(1)$. Therefore, we choose in this scan the values $\beta = 7.32,\, 7.33,\, 7.335,\, 7.34,\,  7.345,\, 7.35$. 

By measuring the ensemble average of the absolute value of the Polyakov loop, as well as  the average plaquette, we determine that the range of values of $\beta$ of interest is $7.33 < \beta < 7.35$. With this information at hand, by  inspecting the histogram of measured average plaquette values, we determine the required plaquette range of interest to be $(u_p)_\mathrm{min} = 0.565 < u_p < (u_p)_\mathrm{max} = 0.585$. As the volume increases, we expect the plaquette distribution to become more constrained, and therefore for larger volumes we restrict this range to $(u_p)_\mathrm{min} = 0.568 < u_p < (u_p)_\mathrm{max} = 0.576$.

We list in Table~\ref{tab:lat_surf1} the parameters used in each of our LLR calculations for the $Sp(4)$ lattice theory,
with $N_t=4$.
On the smaller lattices,  we choose two different configurations for the LLR algorithm, by changing the number of subintervals intervals, $N_{\rm int}$, and the total range of plaquette, $u_p$, considered. 
We do so  to be able to demonstrate  that the calculations are performed in the asymptotic regime of the $\Delta_{u_p} \to 0$ extrapolation.  More details and discussions are given in Appendix~\ref{App:DELim}. As in Ref.~\cite{Lucini:2023irm}, we extrapolate the results linearly in $\Delta_{u_p}^2$, to approach the $\Delta_{u_p} \to 0$ limit. For  $N_s = 48$, due to computational resources we have results for a single interval size. 
For $N_t=5$, we list some of the available information in Appendix~\ref{App:Nt5}---
see Table~\ref{tab:lat_surf2}---but a large scale, high precision analysis is postponed to future publications.

An estimate for $a_n(u_p)$, used as initial condition for the LLR algorithm, is obtained as follows.
We focus at first only on lattices with $\tilde V / a^4 = 4 \times 20^3$.
We use standard importance sampling methods to compute $\langle u_p \rangle_\beta$, for  the six values of $\beta$ listed earlier on. We then fit the result with a cubic polynomial, and invert the outcome (locally) to obtain a function, $\beta(u_p)$. For each lattice volume, 
and each choice of plaquette intervals, we then identify
 the middle point, $(u_p)_n$, of the $n=1,\,\cdots,\,N_{\rm int}$ subintervals.
The corresponding values of the fitted $\beta((u_p)_n)$ are finally used as to compute a first (rough)
 estimate of the coefficients, $\{a_n^{(0)} \}_{n=1}^{N_{\rm int}}$.

As described in Appendix~\ref{App:LLR}---to which we refer for more details---the LLR algorithm takes the aforementioned initial value of $\{a_n^{(0)} \}_{n=1}^{N_{\rm int}}$,
and improves them iteratively, by applying at first the  Newton-Raphson (NR) method, followed by the
Robbins-Monro (RM) method.
By monitoring the NR trajectories of the coefficients, we find that $\bar{m}=10$ iterations of the NR method evolve the coefficients sufficiently close to the true value as to be able to swap to the RM method. We choose the total number of RM iterations by balancing considerations such as computing time constraints and  accuracy goals of this study. For the smallest lattices, we carry out $\tilde{m}=300$ RM iterations,  reduced to $\tilde{m}=100$ for the largest lattice volumes available. In all cases, we verify that the result is converging to the same value at approximately the asymptotic scaling for RM iterations, $1/\sqrt{m}$.


\begin{figure}[t!]
\centering
\includegraphics[width=0.45\textwidth]{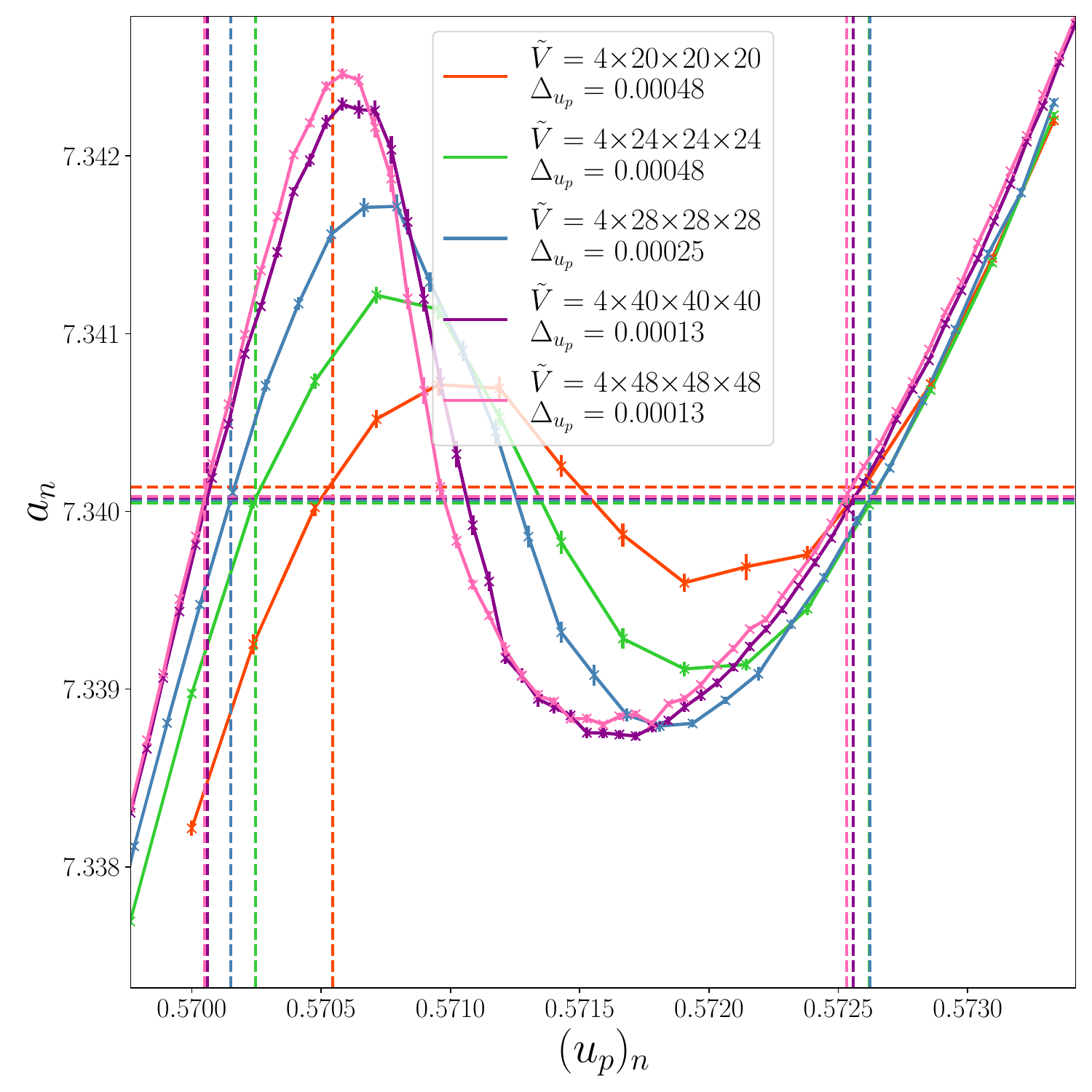}
\caption{\label{fig:an_En_Nt4} Measured coefficients, $a_n$, as a function of the plaquette value at the centre of the energy subinterval, $(u_p)_n$, for the $Sp(4)$ theory, for lattices with temporal size of $N_t=4$, and spatial sizes $N_s = 20,\, 24,\, 28,\, 40,\, 48$ (color coded). The size of the subinterval sizes, for each of the spatial volumes, is given by $\Delta_{u_p} = 0.00048,\, 0.00048,\, 0.00025,\, 0.00013,\, 0.00013$, respectively. The plot displays a portion of the available data, highlighting  the critical region, in which the  noninvertibility of $a_n((u_p)_n)$ is manifest. The horizontal dashed lines shows the critical coupling, $a_n = \beta_{CV}(f)$, 
estimated in the study of the reduced free energy density, $f$. The vertical lines show the corresponding plaquette values.    }
\end{figure}

\begin{figure}[t!]
\centering
\includegraphics[width=0.45\textwidth]{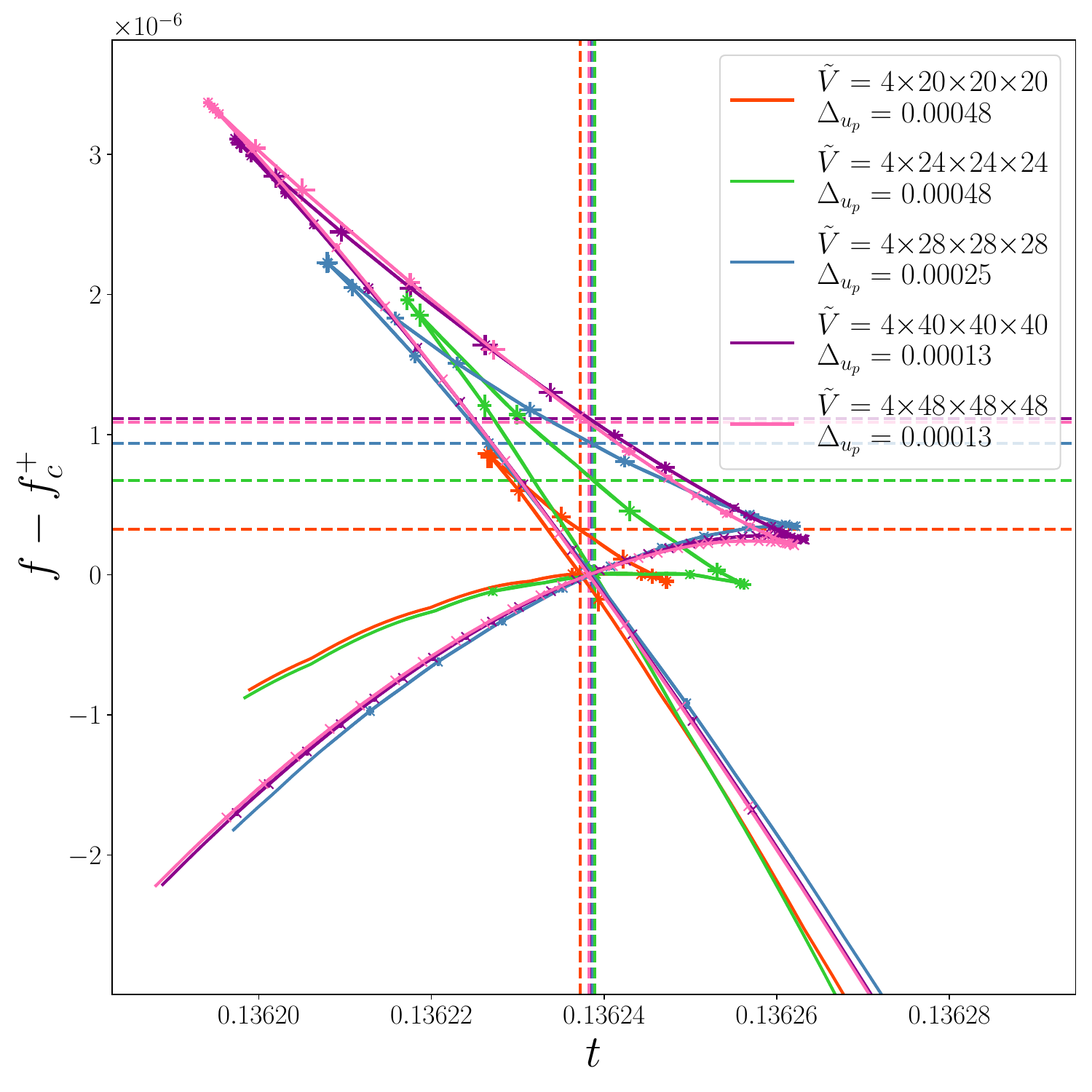}
\caption{\label{fig:t_F_Nt4} The reduced free energy density, $f$, plotted as a function of the microcanonical temperature, $t=1/a_n$, for lattices with temporal size $N_t$ = 4, spatial sizes $N_s= 20,\, 24,\, 28,\, 40,\, 48$ (color coded). The subintervals have sizes $\Delta_{u_p} = 0.00048,\, 0.00048,\, 0.00025,\, 0.00013,\, 0.00013$, respectively. To ensure that the (reduced) free energy difference is directly comparable between lattice sizes, a constant term, $f_c^+$, has been subtracted from $f$. This is defined so that all reduced free energies vanish at the transition, the point in which the two metastable branches cross. The vertical lines show the crossing points between metastable branches, which identifyies the critical temperature. The horizontal lines denote the value of the free energy density along the unstable branch, evaluated  at criticality.}
\end{figure}

In Fig.~\ref{fig:an_En_Nt4}, we display  the final values of the coefficients, $a_n$, obtained with the numerical algorithm, as a function of the plaquette value measured at the centre of each subinterval, $(u_p)_n$. The figure shows the results obtained with the smallest choice of  subinterval size, $\Delta_{u_p}$, for each of the available spatial volumes. The plot shows only a detail of  the critical region, signalled by  the noninvertibility of $a_n(u_p)$. This is a global manifestation of the  metastable dynamics characteristic of a first order transition. 
As the lattice volume increases, the  region of $u_p$ for which $a_n(u_p)$ is not invertible increases in extent, 
and its manifestation sharpens. The coefficients $a_n$ must be volume independent~\cite{Langfeld:2015fua}, hence these are systematic, finite-volume effects that degrades the quality of the measurements, and must be minimised.  For particularly small volumes, one might even miss the noninvertibility signal,
which would become fainter and more difficult to measure.
By contrast, as the figure shows, with our choices of parameters, subintervals, and analysis strategy, the results converge well towards the thermodynamic limit, $N_s\rightarrow +\infty$.   

\begin{figure*}[th]
\subfloat[$\tilde V/a^4 = 4\times 20^3$, $\Delta_{u_p} = 0.00048$\label{fig:Pbup_420}]
			{\includegraphics[width=0.29\textwidth]{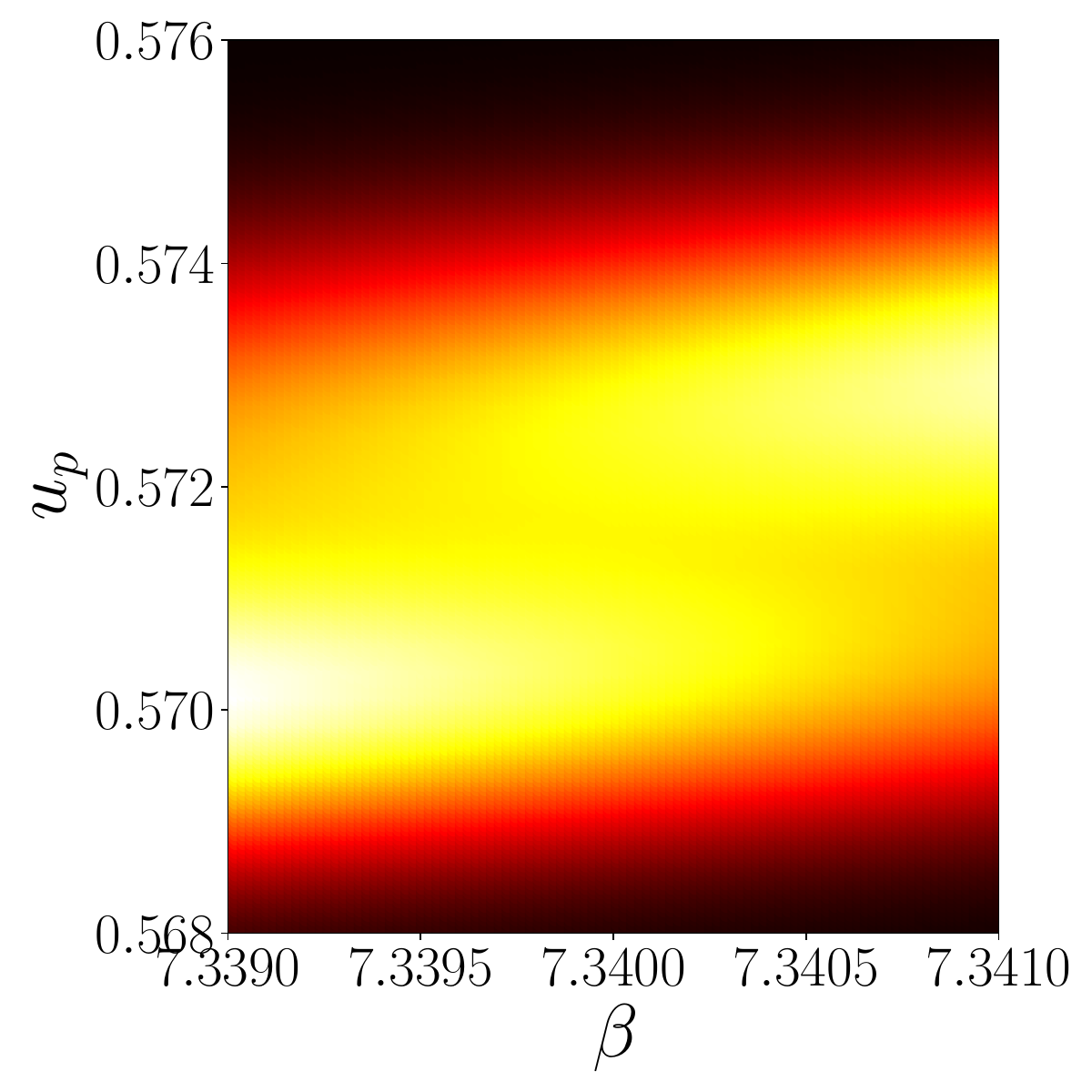}}\quad
\subfloat[$\tilde V/a^4 = 4\times 24^3$, $\Delta_{u_p} = 0.00048$\label{fig:Pbup_424}]	
			{\includegraphics[width=0.29\textwidth]{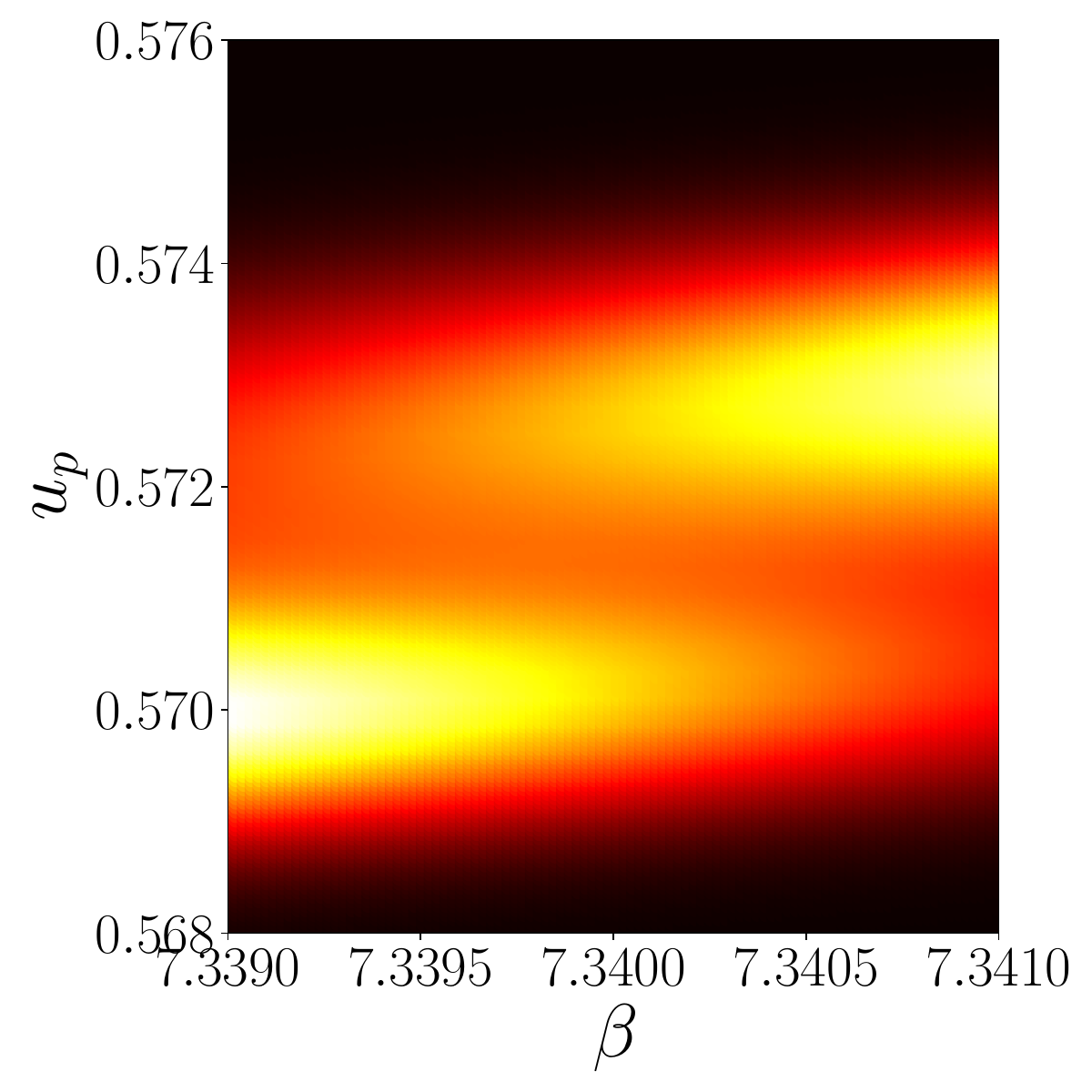}	}	
\subfloat[$\tilde V/a^4 = 4\times 28^3$, $\Delta_{u_p} = 0.00025$\label{fig:Pbup_428}]
		{	\includegraphics[width=0.29\textwidth]{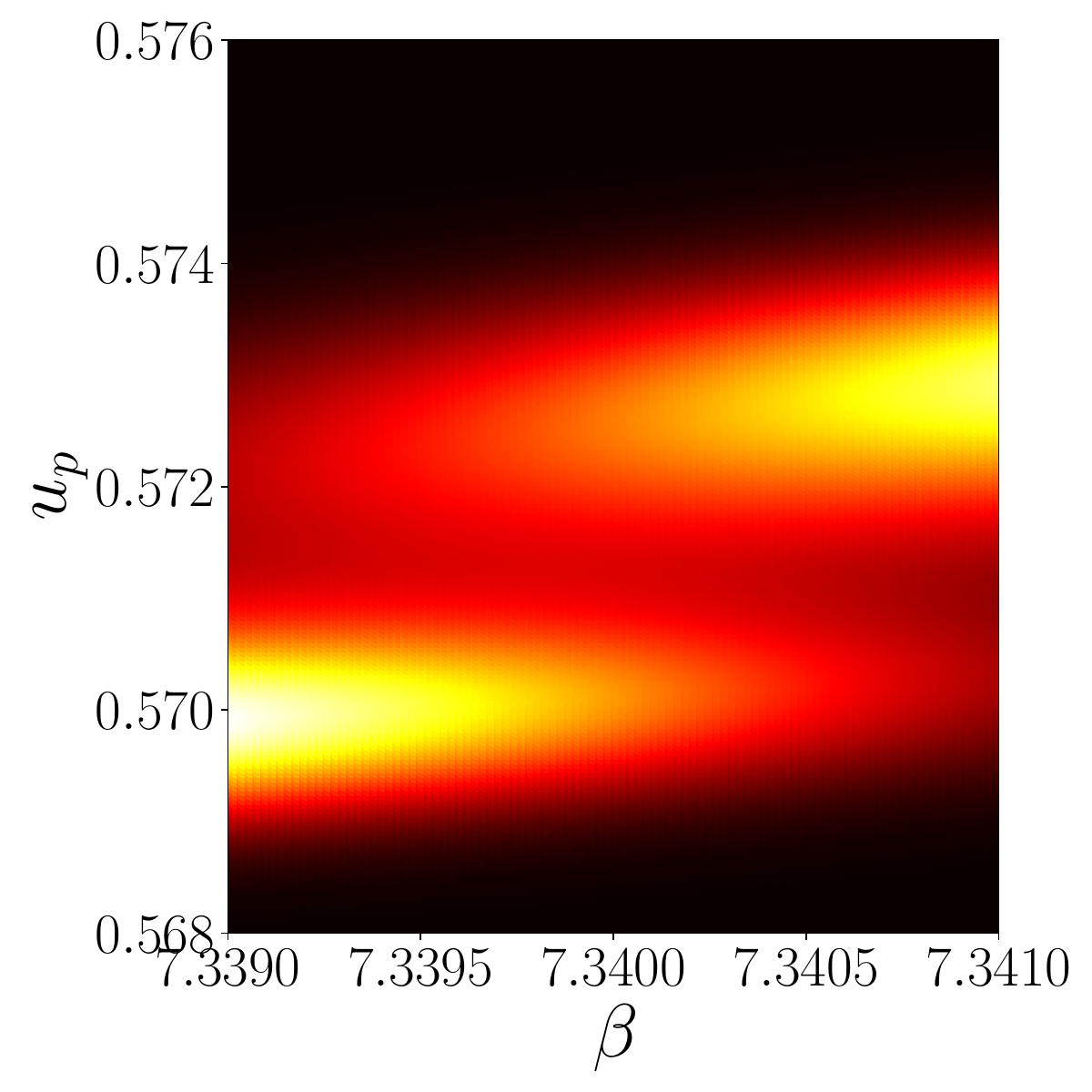}}
			\quad
\subfloat[$\tilde V/a^4 = 4\times 40^3$, $\Delta_{u_p} = 0.00013$\label{fig:Pbup_440}]
			{\includegraphics[width=0.29\textwidth]{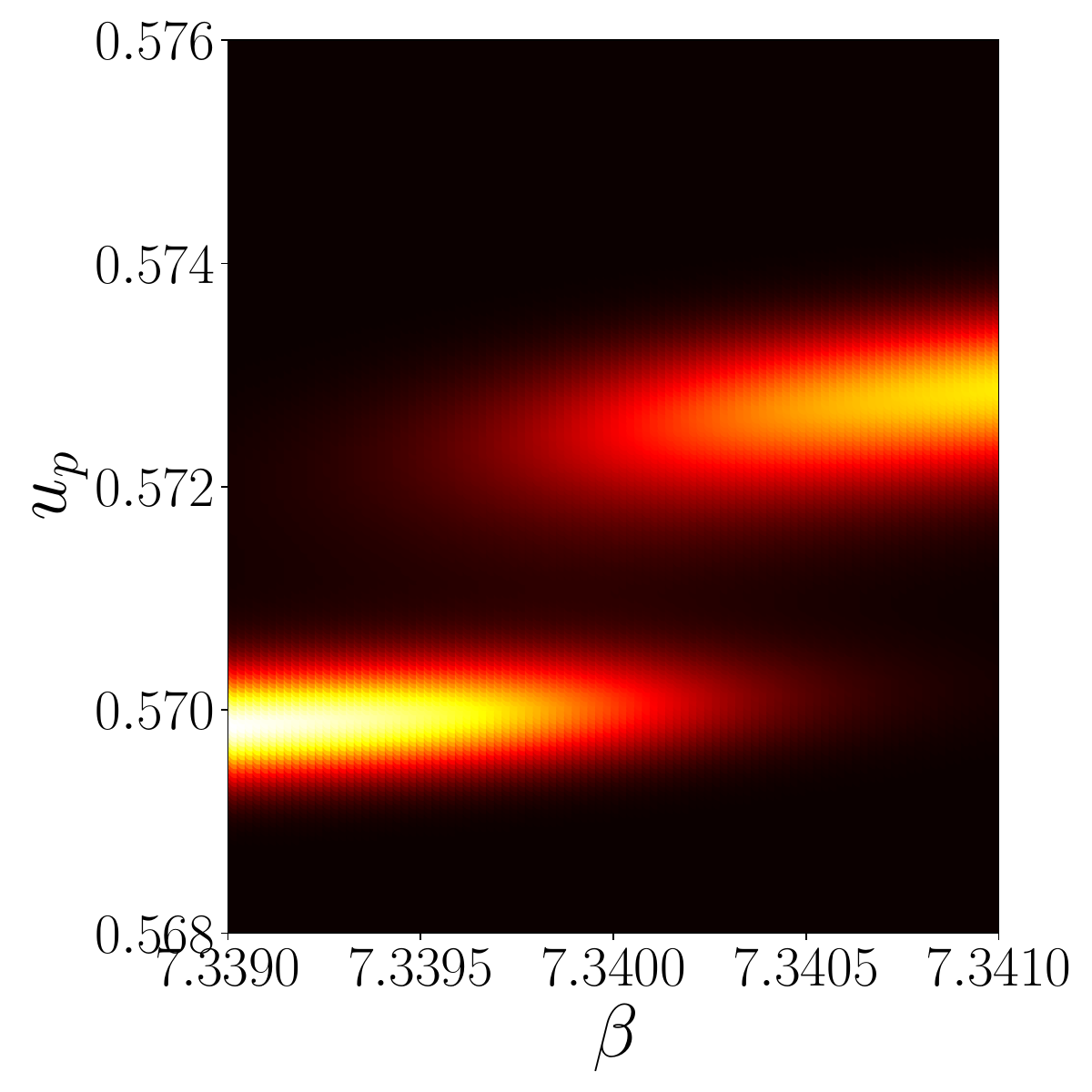}	}	
\subfloat[$\tilde V/a^4 = 4\times 48^3$, $\Delta_{u_p} = 0.00013$\label{fig:Pbup_448}]
			{\includegraphics[width=0.29\textwidth]{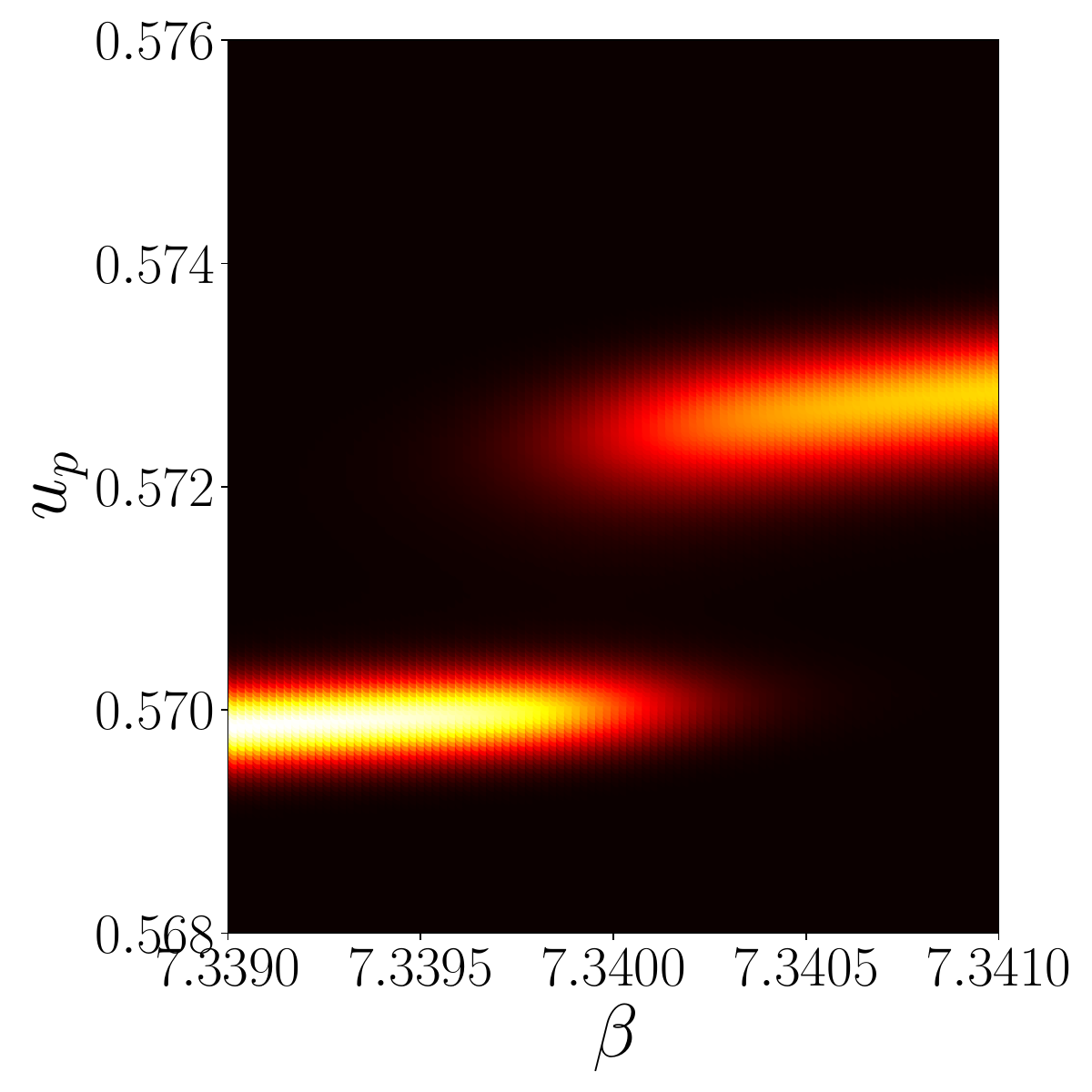}}		
	
        \hfill
	\caption{The plaquette probability distribution, $P_\beta(u_p)$, displayed as a colour map in the plane of coupling, $\beta$, and plaquette value, $u_p$. Darker colours represent lower $P_\beta(u_p)$ values. The temporal lattice size is $N_t = 4$, with different spatial sizes, $N_s = 20,\, 24,\, 28,\, 40,\, 48$, and  with subinterval sizes $\Delta_{u_p} = 0.00048,\, 0.00048,\, 0.00025,\, 0.00013,\, 0.00013$, respectively. The plaquette range displayed in all plots is  $ 0.568<u_p < 0.576$, 
	with  coupling in the range  $7.339 < \beta <  7.341$. }
	\label{fig::Pbup_4}
\end{figure*}

By using the coefficients $a_n$, we compute the thermodynamic potentials Eq.~(\ref{eq:Therm_pot}). By piecewise inverting $a_n(u_p)$, and using Eq.~(\ref{eq:Red_FE}), to find $E(t)$, $s(t)$ and therefore the (reduced) free-energy, $f(t)$, which we then show in Fig.~\ref{fig:t_F_Nt4}, for the same choices of volume and subintervals as in Fig.~\ref{fig:an_En_Nt4}. 
In doing so, we set the additive integration constant,  $\Sigma$, appearing in the entropy, $s$, to be equal to the average entropy across all intervals.\footnote{This choice leads to the visible difference in rotation of the tail of the diagram for $N_s = 20$ and $N_s = 24$, which is due to the difference in the  range of plaquette used in the calculation of $\Sigma$.}  The points on the plot show the values for the centre of each interval, with errors in both the value of the reduced free energy and the (microcanonical) temperature, calculated by bootstrapping over $n_R$ repeats of the LLR algorithm. The lines are found by piece-wise linearly interpolating in the interval between the available points for $a_n((u_p)_n)$.

This is possibly the most important, certainly the most beautiful result of this study, displayed in Fig.~\ref{fig:t_F_Nt4}.
The free energy density, $f$,  displays the swallow-tail structure that is characteristic of a first-order transition. Away from the critical region, $f$ is single valued, corresponding to plaquette values, or energies, for which only one single phase exists. In the critical region, there are two co-existing metastable branches, each corresponding to pure phase states, and an unstable, mixed phase branch, having the largest free energy. 
We observe that the region exhibiting metastable dynamics grows with the volume as does the difference 
in free energy between the unstable and metastable branches. Yet, it appears to converge in the thermodynamic limit, $N_s\rightarrow +\infty$.        

From Fig.~\ref{fig:t_F_Nt4},  the reciprocal of the (micro-canonical) temperature, $t$, at which the two meta-stable branches cross, or equivalently the value of the coefficient, $a_n$, for which this happens,
  provides a measurement of the critical coupling of the system, $\beta_{CV}(f) = 1/t_c$.
At this value, the plaquette distribution has a double Gaussian structure with two peaks of equal height.
Numerically, we determine the critical point by finding the value of $t$ which minimises the difference in free-energy between the two meta-stable branches. 
Uncertainties, as elsewhere, are computed by bootstrapping over the results for the $n_R$ repeats. 
Having determined the critical coupling, we compute the
 reduced free energy of the metastable branch at the critical point, $f_c^+$.
As we are only interested in the free-energy difference between the micro-states at the same temperature, and we want to compare different calculations across volumes, we remove a second additive integration constant, in $f$, by subtracting the  $f_c^+$.
The horizontal lines in Fig.~\ref{fig:t_F_Nt4} show the value of the reduced free-energy on the unstable branch at the critical point. 

\begin{figure}[t!]
\centering
\includegraphics[width=0.45\textwidth]{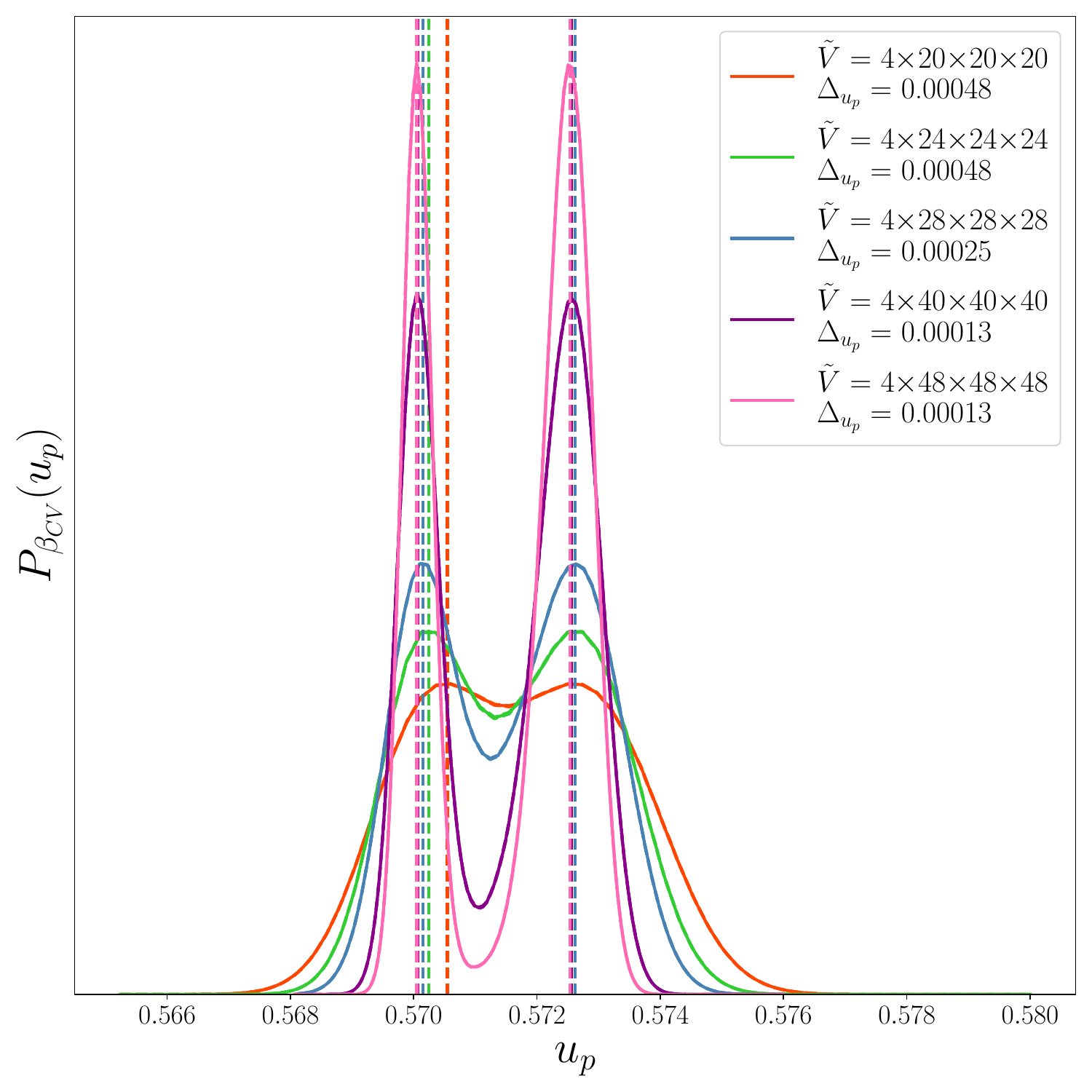}
\caption{\label{fig:PbE_Nt4} The plaquette probability distribution, $P_\beta(u_p)$, computed  at the critical point, $\beta=\beta_{CV}(f)$, exhibiting a double peak structure with peaks of equal height. The plot shows results for lattice systems with temporal extent $N_t = 4$, spatial sizes, $N_s = 20,\, 24,\, 28,\, 40,\, 48$, (color coded), and  subinterval sizes $\Delta_{u_p} = 0.00048,\, 0.00048,\, 0.00025,\, 0.00013,\, 0.00013$, respectively. The vertical dashed lines show the location of the peaks of the distribution.}
\end{figure}

From the properties of the reduced free energy density we can also compute the other two main thermodynamic observables characterising the transition. For the interface tension, the logarithmic term in Eq.~(\ref{eq:Interface0}), which determines $\tilde{\sigma}_{cd}$ via Eq.~(\ref{eq:Interface}),
is related to the difference, $\Delta f$,  in the (reduced) free energy between the unstable and  metastable branches at the critical point, $\log(P_\mathrm{max} / P_\mathrm{min} ) = (\tilde V / a^4 ) \Delta f / t_c$. 
For the latent heat, we need to measure the discontinuity in the plaquette at the critical point, $\Delta \langle u_p \rangle_{\beta_{CV}}$. It can be extracted from the function $u_p(a_n)$, presented in Fig.~\ref{fig:an_En_Nt4}. In this plot the horizontal dashed line shows $a_n = \beta_{CV}(f)$, the vertical lines show the plaquette values which intercept this line, $u_p(a_n = \beta_{CV}(f))$, along the metastable branches. These two values correspond to the peaks of the plaquette distribution at the critical point. Their difference is the jump (discontinuity) in the plaquette.

We can compute the distribution probability of the plaquette, $P_\beta(u_p)$, for any value of the coupling in the range for which we computed the coefficients $a_n$, by
using Eq.~(\ref{eqn:PlaqDistribution}), with $\rho$ replaced by $\tilde{\rho}$, the latter  being determined by the measured values of $a_n$.
 The colour maps in Fig.~\ref{fig::Pbup_4} show the resulting plaquette distribution, as reconstructed from the LLR algorithm, in the critical region between $ 7.339<\beta <7.341$. 
 Starting from the lowest value of the coupling, $\beta$, each plot displays the presence of a single peak, corresponding to a single available phase for the system.
The peak decreases in size and spreads out when increasing $\beta$. A second peak starts to appear and grows until it has equal height to the first peak, at the critical coupling (towards the middle of the plots). Then, as the coupling is increased further, the new peak from the high temperature phase starts to dominates the system while the original peak starts to fade. Ultimately, at the right-hand of the range of $\beta$ displayed, only the distribution due to the new phase is still visible. As the spatial volume grows, the width of the peaks decreases and the height of the peak grows. It is expected they will become Dirac delta distributions in the thermodynamic limit.

\begin{figure}[t!]
\centering
\includegraphics[width=0.45\textwidth]{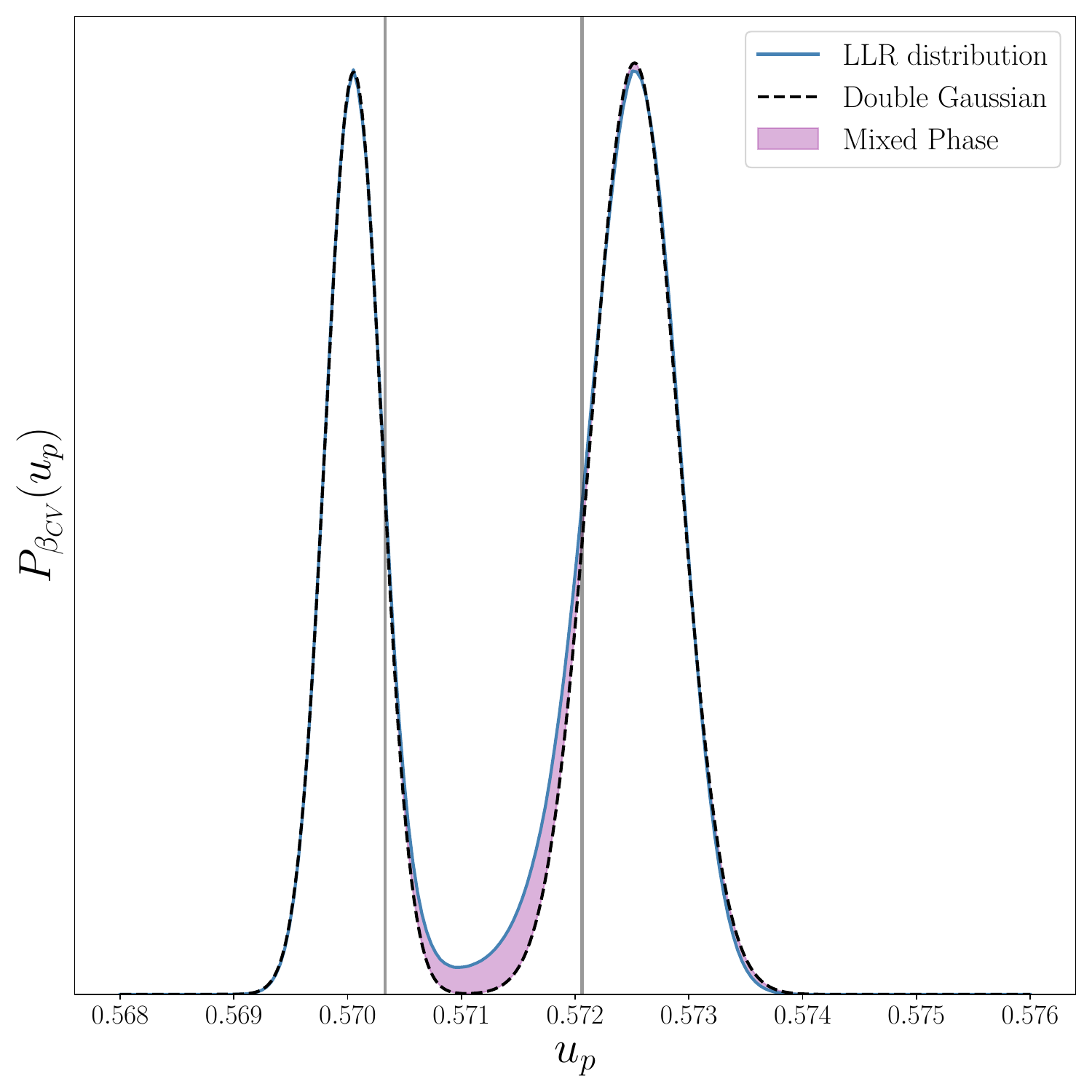}
\caption{\label{fig:Interface} The plaquette probability distribution, $P_\beta(u_p)$, computed at the critical point, $\beta=\beta_{CV}(f)$, and exhibiting a double peak structure with peaks of equal height. The plot shows results for a lattice systems of volume $\tilde V / a^4 = 4\times 48^3$, with interval sizes $\Delta_{u_p} = 0.00013$. The black dashed line shows a fitted double Gaussian function to the observed plaquette distribution reconstructed from the LLR results, in blue. The double Gaussian is fitted to the region outside the grey verticalal lines, therefore excluding the central dip. The difference between the observed plaquette distribution and the fitted double Gaussian is highlighted in pink.}
\end{figure}

Looking only at Fig.~\ref{fig::Pbup_4}, one might incorrectly infer that the probability distribution vanishes in the  area between the peaks. However, in  Fig.~\ref{fig:PbE_Nt4}
one can see that there remains a finite probability  between the peaks, at criticality, for all spatial volumes.
As the volume increases, the peak of the distribution grow while their width decreases. While for the smallest volumes the whole distribution can be approximated as a double Gaussian, 
this approximation fails at larger volumes,
as the region between the peaks displays a finite plateau. 

Figure~\ref{fig:Interface} demonstrates this
phenomenon, for the largest volume available, and for subintervals with width $\Delta_{u_p}=  0.00013$. 
We fit a double Gaussian 
 to the LLR results, but exclude from the fit
 the central region, between vertical lines. The double Gaussian approximation holds well in the external region, included in the fit.  However, in the internal region the double Gaussian vanishes, while the LLR distribution converges to a constant. 
 We expect the primary reason for this difference to be due to the presence of mixed phase states, and the interface tension between them.

\begin{figure}[t!]
\centering
\includegraphics[width=0.45\textwidth]{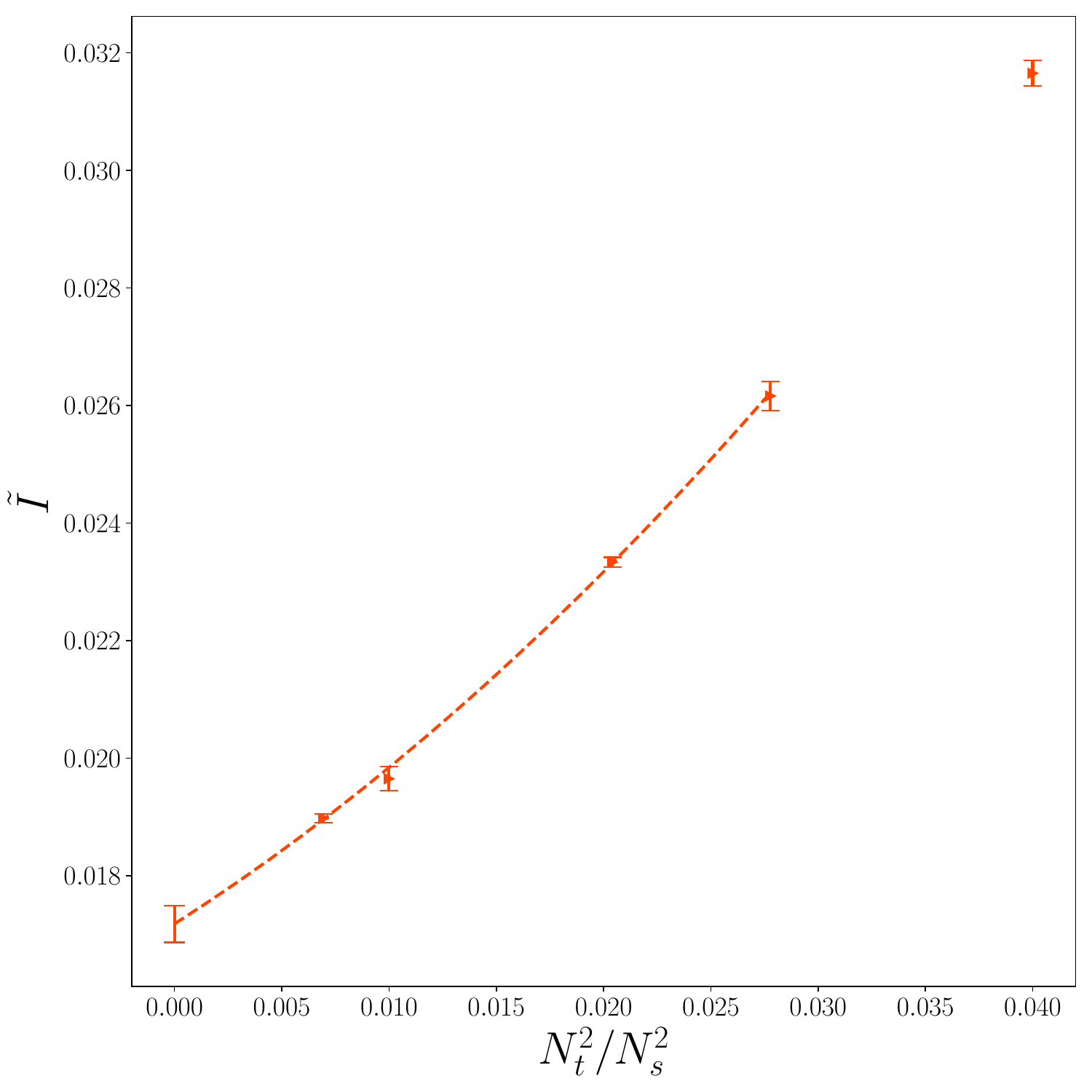}
\caption{\label{fig:I_Nslim_Nt4} The interface term, $\tilde I$, defined in Eq.~(\ref{eq:Interface0}), plotted against the square of the inverse of the aspect ratio, $N_t^2 / N_s^2$, for lattices with temporal extent $N_t$ = 4, and spatial sizes $N_s = 20,\, 24,\, 28,\, 40,\, 48$. For each spatial volume, we show the result of the extrapolation to the limit of vanishing subinterval size---see Appendix~\ref{App:DELim}. The dashed line shows a 
 polynomial fit in $N_t^2 / N_s^2$ of the points with the four largest spatial volumes. The reduced chi square is $\chi^2/N_{\rm d.o.f.} = 1.0$. We also display the extrapolated $\lim_{N_t/N_s\rightarrow 0}\tilde I = 0.01718(32)(1)$.}
\end{figure}

To quantify the effect of the interface terms, we compute the quantity $\tilde I$, defined in Eq.~(\ref{eq:Interface0}), which relates to the interface tension, $\tilde{\sigma}_{cd}$, and therefore the probability of tunnelling between phases. We compute the logarithmic term, $\log(P_\mathrm{max} / P_\mathrm{min})$, in terms of the change in free-energy, as explained earlier. Errors are, again, computed by bootstrapping the results over the values found for the $n_R$ independent repeats. To remove finite interval systematic effects, we extrapolate the results to the limit of vanishing interval size, for each spatial volume.
Details are provided in Appendix~\ref{App:DELim}. The measured values of $\tilde I$ are presented in Fig.~\ref{fig:I_Nslim_Nt4}, as a function of the square of the inverse of the aspect ratio, $(N_t / N_s)^2$. We fit a quadratic polynomial in $(N_t / N_s)^2$  to our numerical results, excluding the smallest volume, $N_s = 20$, and extrapolate to the thermodynamic limit, $(N_t / N_s) \to 0$. The reduced chi-square for the fit is given by $\chi^2/N_{\rm d.o.f.} = 1.0$, and the extrapolation gives $\tilde \sigma_{cd} = \lim_{N_t/N_s\rightarrow 0}\tilde I = 0.01718(32)(1)$\footnote{The second bracket denotes an additional systematic uncertainty coming from the chosen parameterization of the thermodynamic limit, as discussed in Appendix \ref{App:Systematic}.}. As a side remark, we note that  in the case of the two largest available volumes, with $N_s = 40$ and $N_s = 48$, the  term logarithmic in $N_s$ dominates in Eq.~(\ref{eq:Interface0}), and their subtraction is essential. Therefore, it is possible that only the final two points are close enough to the asymptotic regime (the thermodynamic limit). Yet, the overall trend of the data does not display reasons for concern, with current uncertainties.

\begin{figure}[t!]
\centering
\includegraphics[width=0.45\textwidth]{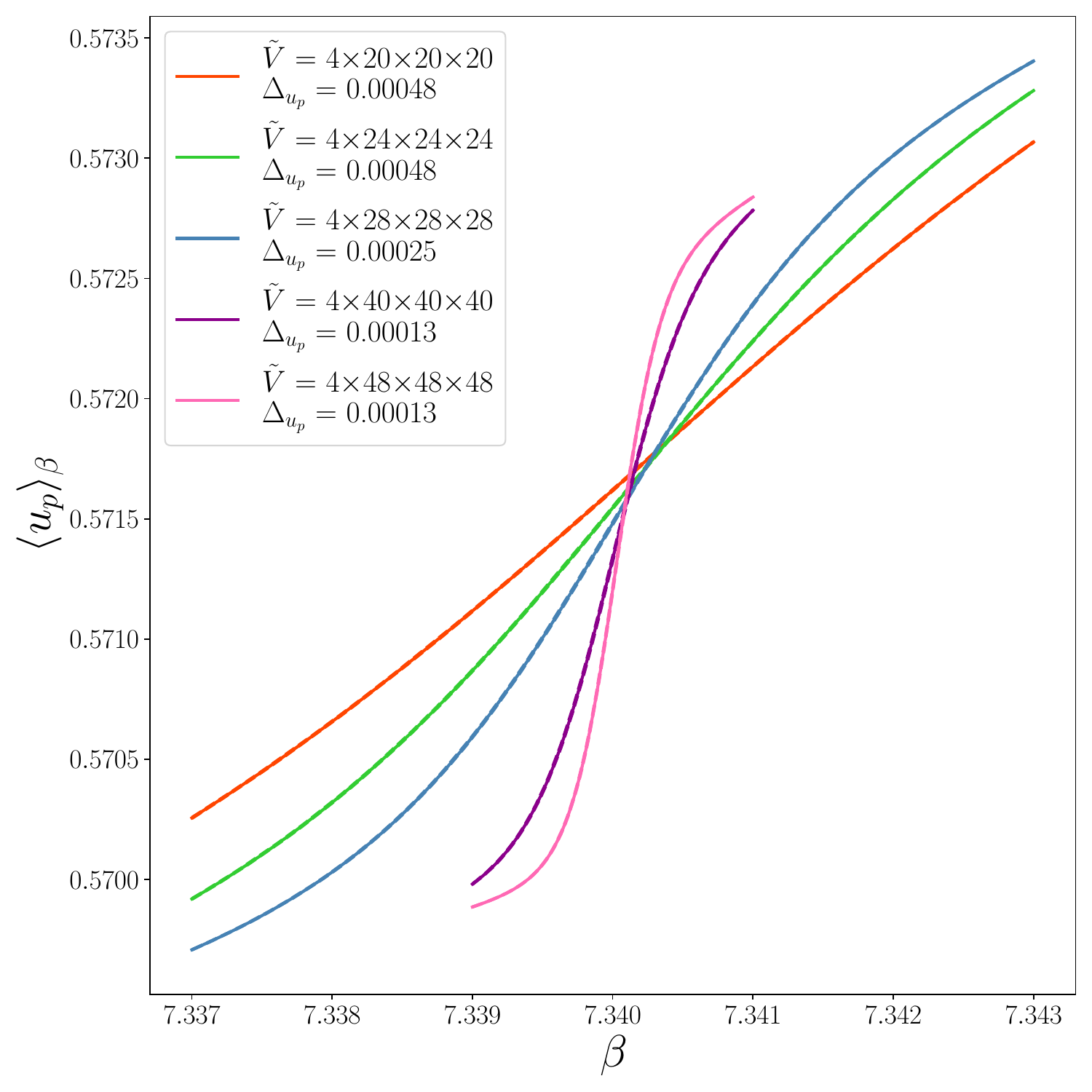}
\caption{\label{fig:u_Nt4} The VEV of the average plaquette, $\langle u_p \rangle_\beta$, plotted as a function of  coupling, $\beta$, for lattices with temporal size, $N_t$ = 4, spatial sizes $N_s = 20,\, 24,\, 28,\, 40,\, 48$ (color coded), and subinterval sizes $\Delta_{u_p} = 0.00048,\, 0.00048,\, 0.00025,\, 0.00013,\, 0.00013$, respectively. 
We chose a grid with 1000 evenly spaced values of $\beta$, in the interval $7.337 <\beta<7.343$, for $N_s= 20, 
\,24$, and $28$, and $7.339<\beta<7.341$, for $N_s = 40$ and $48$.  }
\end{figure}

\begin{figure}[t!]
\centering
\includegraphics[width=0.45\textwidth]{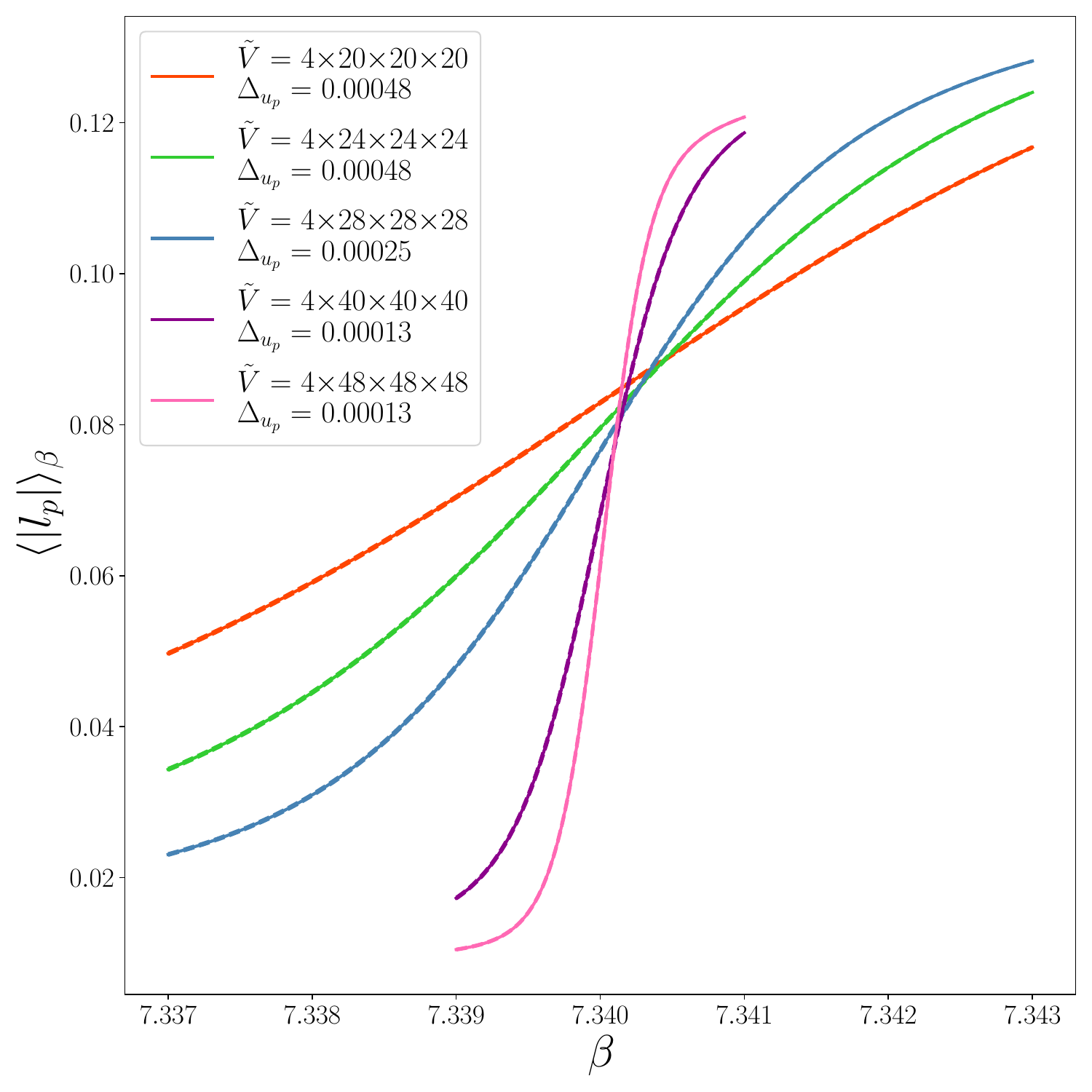}
\caption{\label{fig:lp_Nt4} The VEV of the absolute value of the Polyakov loop, $\langle | l_p | \rangle_\beta$, plotted as a function of  coupling, $\beta$, for lattices with temporal size, $N_t$ = 4, spatial sizes $N_s = 20,\, 24,\, 28,\, 40,\, 48$ (color coded), and subinterval sizes $\Delta_{u_p} = 0.00048,\, 0.00048,\, 0.00025,\, 0.00013,\, 0.00013$, respectively. 
We chose a grid with 100 evenly spaced values of $\beta$, in the interval $7.337 <\beta<7.343$, for $N_s= 20, 
\,24$, and $28$, and $7.339<\beta<7.341$, for $N_s = 40$ and $48$.  }\end{figure}

\begin{figure}[t!]
\centering
\includegraphics[width=0.45\textwidth]{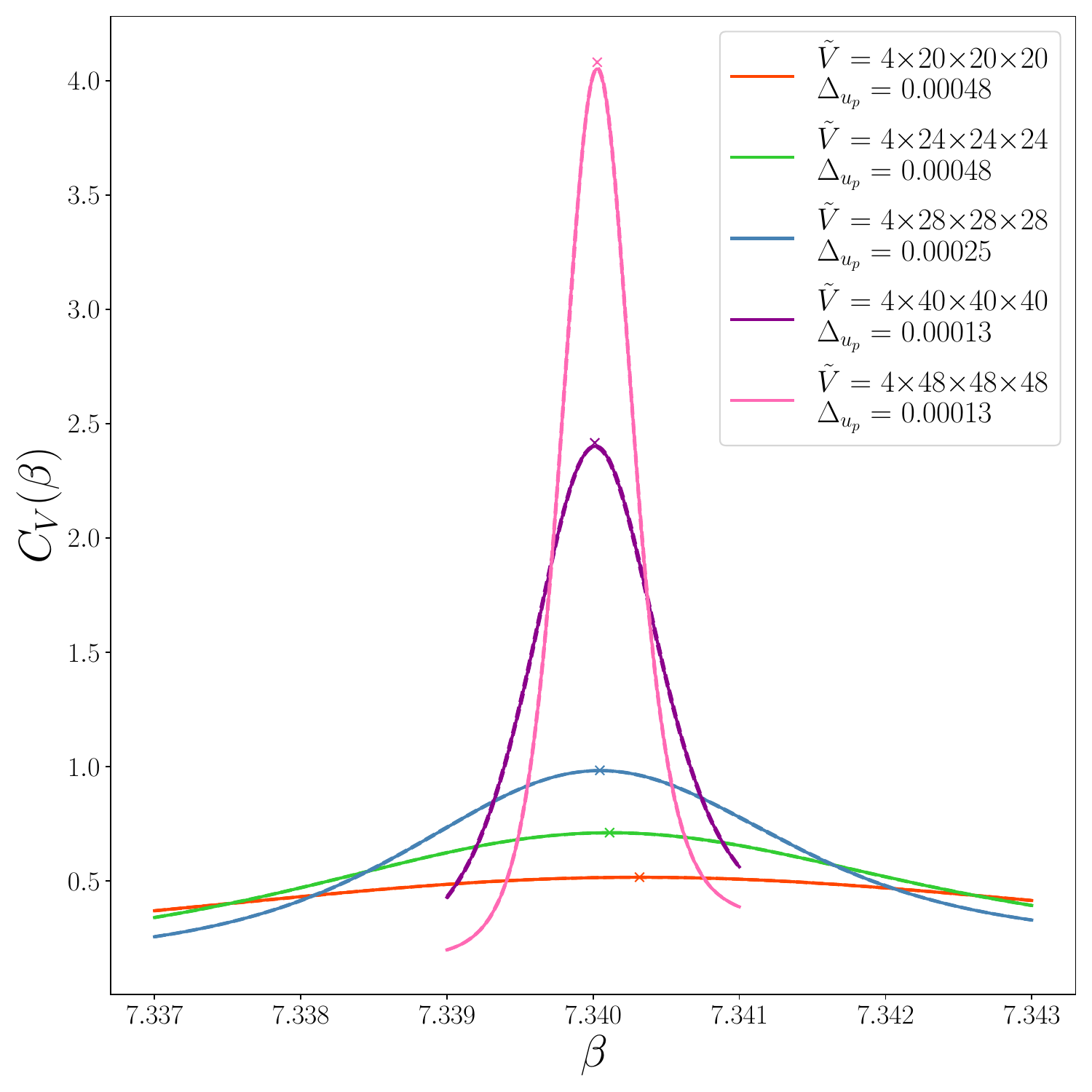}
\caption{\label{fig:Cu_Nt4} The specific heat, $C_V(\beta)$, plotted as a function of  coupling, $\beta$, for lattices with temporal size, $N_t$ = 4, spatial sizes $N_s = 20,\, 24,\, 28,\, 40,\, 48$ (color coded), and subinterval sizes $\Delta_{u_p} = 0.00048,\, 0.00048,\, 0.00025,\, 0.00013,\, 0.00013$, respectively. 
We chose a grid with 1000 evenly spaced values of $\beta$, in the interval $7.337 <\beta<7.343$, for $N_s= 20, 
\,24$, and $28$, and $7.339<\beta<7.341$, for $N_s = 40$ and $48$.  }
\end{figure}

\begin{figure}[t!]
\centering
\includegraphics[width=0.45\textwidth]{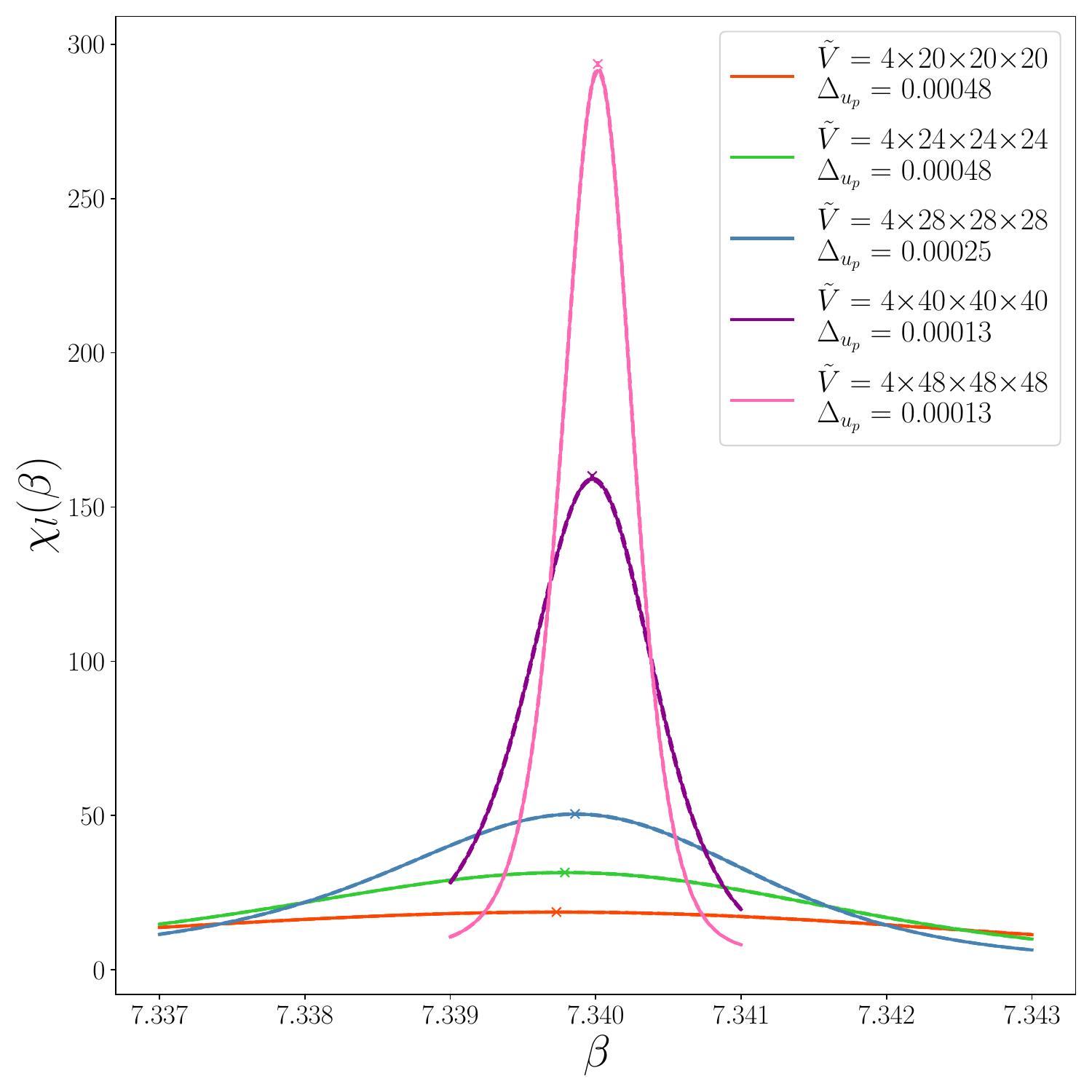}
\caption{\label{fig:Xlp_Nt4} The Polyakov loop susceptibility, $\chi_l(\beta)$, plotted as a function of  coupling, $\beta$, for lattices with temporal size, $N_t$ = 4, spatial sizes $N_s = 20,\, 24,\, 28,\, 40,\, 48$ (color coded), and subinterval sizes $\Delta_{u_p} = 0.00048,\, 0.00048,\, 0.00025,\, 0.00013,\, 0.00013$, respectively. 
We chose a grid with 100 evenly spaced values of $\beta$, in the interval $7.337 <\beta<7.343$, for $N_s= 20, 
\,24$, and $28$, and $7.339<\beta<7.341$, for $N_s = 40$ and $48$.  }
\end{figure}

\begin{figure}[t!]
\centering
\includegraphics[width=0.45\textwidth]{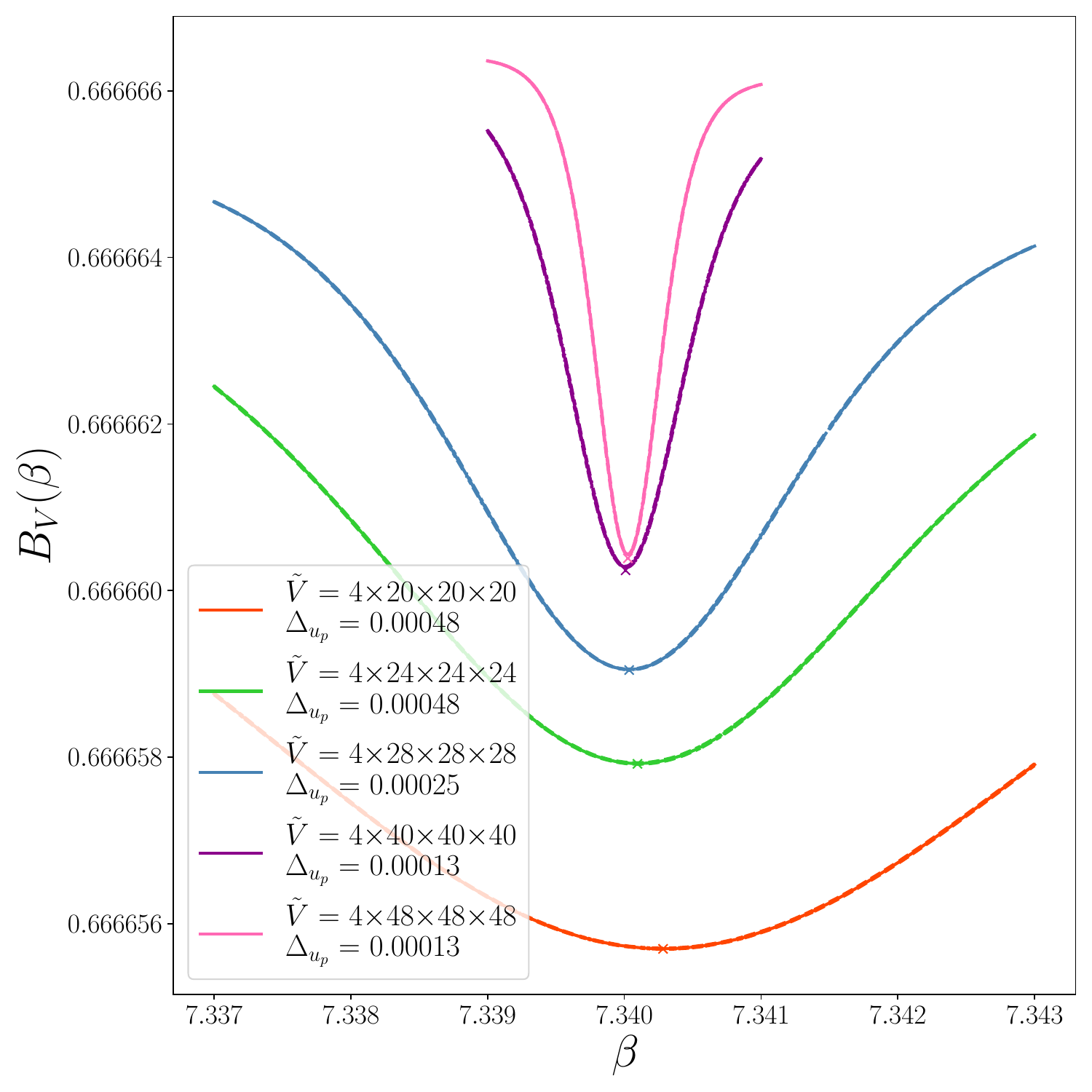}
\caption{\label{fig:Bv_Nt4} The Binder cumulant, $B_V(\beta)$, plotted as a function of  coupling, $\beta$, for lattices with temporal size, $N_t$ = 4, spatial sizes $N_s = 20,\, 24,\, 28,\, 40,\, 48$ (color coded), and subinterval sizes $\Delta_{u_p} = 0.00048,\, 0.00048,\, 0.00025,\, 0.00013,\, 0.00013$, respectively. 
We chose a grid with 1000 evenly spaced values of $\beta$, in the interval $7.337 <\beta<7.343$, for $N_s= 20, 
\,24$, and $28$, and $7.339<\beta<7.341$, for $N_s = 40$ and $48$.  }
\end{figure}

We can reconstruct canonical observables using Eq.~(\ref{eq:vev_obs}), for observables with explicit dependence on the energy, and Eq.~(\ref{eqn:vev_gen}), for more general observables. 
We display some interesting results of this type of  analysis in Figs.~\ref{fig:u_Nt4},~\ref{fig:lp_Nt4},~\ref{fig:Cu_Nt4}, and~\ref{fig:Xlp_Nt4}, and comment about them in the following.

We compute the average plaquette, $\langle u_p \rangle_\beta$, and the absolute value of the Polyakov loop, $\langle | l_p | \rangle_\beta$, for a selection of couplings, $\beta$,  around the critical point, and display the results  in Figs.~\ref{fig:u_Nt4} and~\ref{fig:lp_Nt4}. As the spatial volume increases, the results for these observables approximate the discrete jump one expects in a first-order phase transition. This transition is quantified by the specific heat, $C_V(\beta)$, for the average plaquette, and the Polyakov loop susceptibility, $\chi_l(\beta)$. We display our results for these quantities  in Figs.~\ref{fig:Cu_Nt4} and~\ref{fig:Xlp_Nt4}, respectively. In proximity of a first-order phase transition, we expect the peak of the specific heat, $C_V^\mathrm{(max)}$, and Polyakov loop susceptibility, $\chi_l^\mathrm{(max)}$ to scale with the spatial volume of the system. In both cases, the peak grows and the width decreases with the volume. We measure the values at the peak, $C_V^\mathrm{(max)}$, and $\chi_l^\mathrm{(max)}$, as well as the critical couplings, $\beta_{CV}(C_V)$ and $\beta_{CV}(\chi_l)$, by taking the maximum observed value in our scan of $\beta$ values. We estimate the errors by bootstrapping the results for the $n_R$ repeats.

Another important quantity in the study of first order phase transitions is the Binder cumulant, $B_V(\beta)$, defined in Eq.~(\ref{eq:binder}). At the critical point, this quantity exhibits a minimum, $B_V^\mathrm{(min)}$, in correspondence of the  critical coupling, $\beta_{CV}(B_V)$, which we compute by taking the minimum observed value from a scan of $\beta$ values.  Figure~\ref{fig:Bv_Nt4} shows this quantity, measured in a neighbourhood of the phase transition, for several choices of spatial volume. Although the dip reduces as the volume grows, it stays finite, and appears to converge.

\begin{figure}[t!]
\centering
\includegraphics[width=0.45\textwidth]{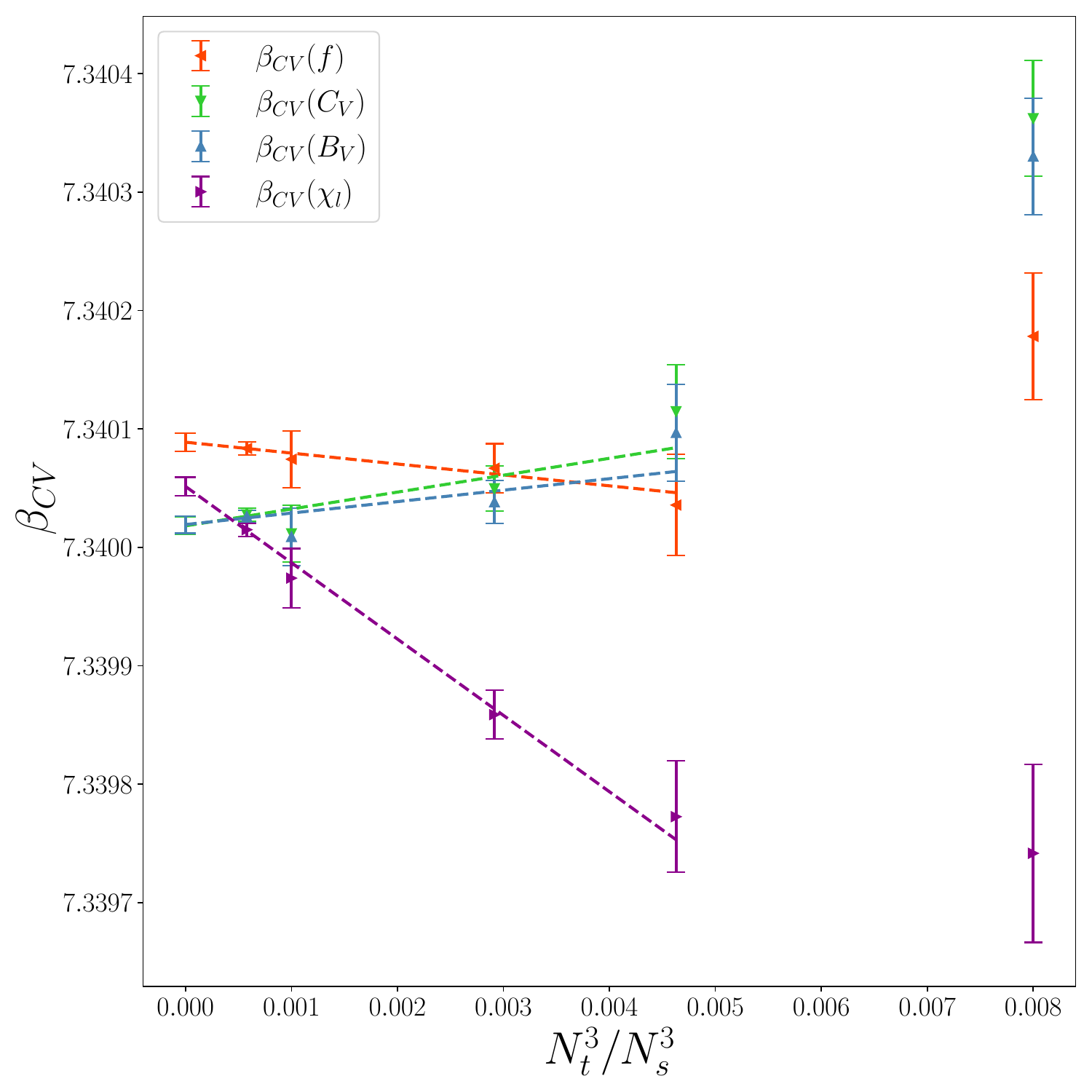}
\caption{\label{fig:Bc_Nslim_Nt4} Extrapolation towards the thermodynamic limit of the  critical coupling
defined in four alternative ways. The first definition is based on the crossing between the two metastable branches  in the study of $f$, as displayed in Fig.~\ref{fig:t_F_Nt4}, and denoted $\beta_{CV}(f)$.  The other three definitions
involve locating  the extrema of the specific heat, $\beta_{CV}(C_V)$, the Binder cumulant, $\beta_{CV}(B_V)$, and the Polyakov loop susceptibility, $\beta_{CV}(\chi_l)$. We show the four measurements  in different colours,
for each available different volume, or the inverse of the aspect ratio $N_t^3 / N_s^3$, at fixed temporal size, $N_t$ = 4.
The spatial sizes are  $N_s = 20,\, 24,\, 28,\, 40,\, 48$.  For each spatial volume, we use measurements extrapolated to  the limit of vanishing interval size---see Appendix~\ref{App:DELim}. 
Linear fits of the four largest spatial sizes is shown by the dashed lines, and the extrapolations to the $N_s \to \infty$ are displayed at the left of the plot.}
\end{figure}

\begin{figure}[t!]
\centering
\includegraphics[width=0.45\textwidth]{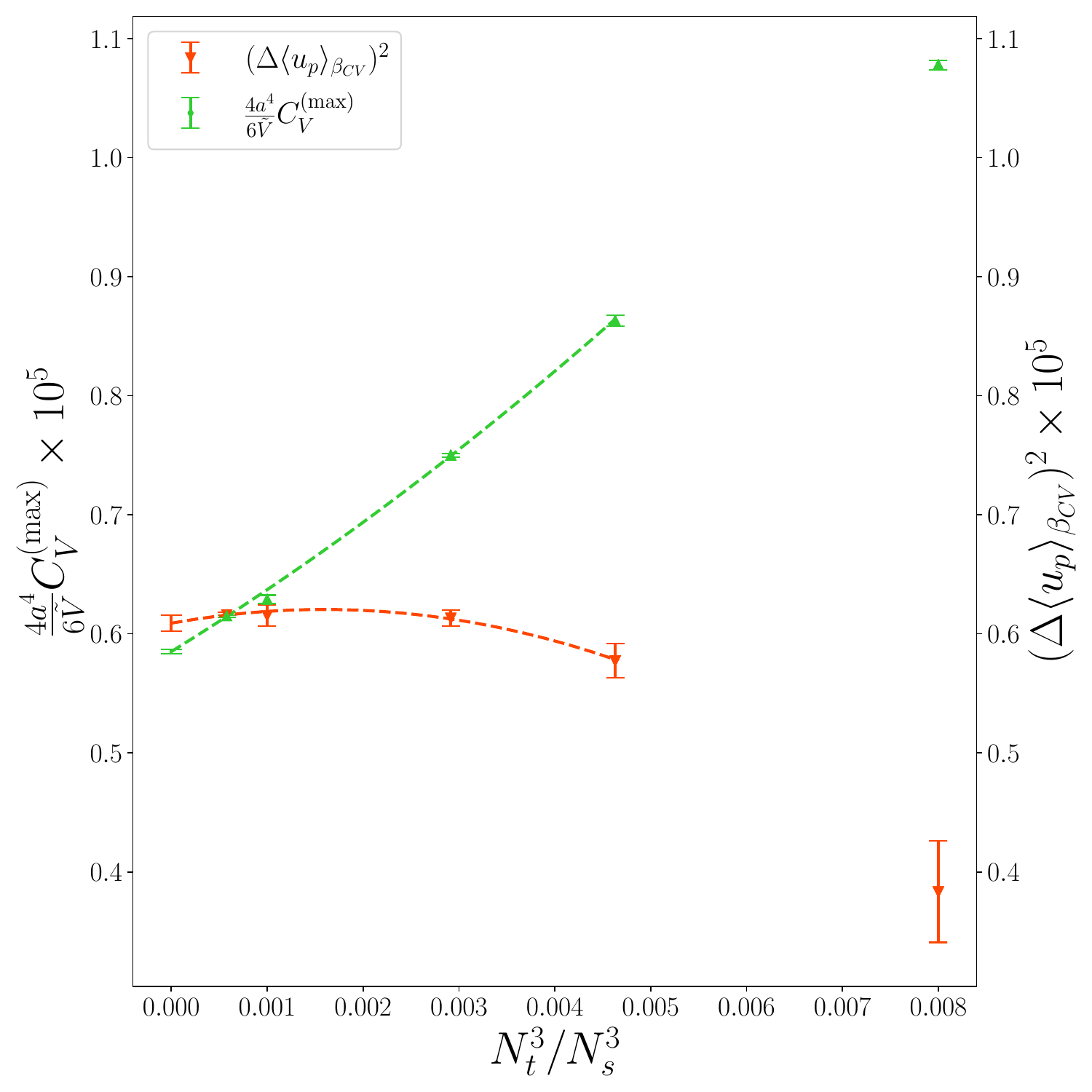}
\caption{\label{fig:DE2_Nslim_Nt4} The square of the plaquette discontinuity measured  at the critical point, 
$(\Delta \langle u_p \rangle_{\beta_{CV}})^2$  (red), and the peak of the specific heat, normalised by the volume, $\frac{4a^4}{6\tilde V}C_V^\mathrm{(max)}$  (green), as a function of the cube of the inverse of the aspect ratio, $N_t^3 / N_s^3$. 
The two quantities as expected to agree in the thermodynamic limit---see Eq.~(\ref{eq:spec_to_latheat}).
The  lattices have temporal size $N_t$ = 4, and spatial sizes $N_s = 20,\, 24,\, 28,\, 40,\, 48$.  For each spatial volume we extrapolate to the limit of vanishing subinterval size---see Appendix~\ref{App:DELim}. The dashed lines show  the results of quadratic polynomial fits of  the four largest spatial volumes. The thermodynamic limits,  $N_s \to \infty$, are displayed at the far left of the plot.}
\end{figure}

\begin{figure}[t!]
\centering
\includegraphics[width=0.45\textwidth]{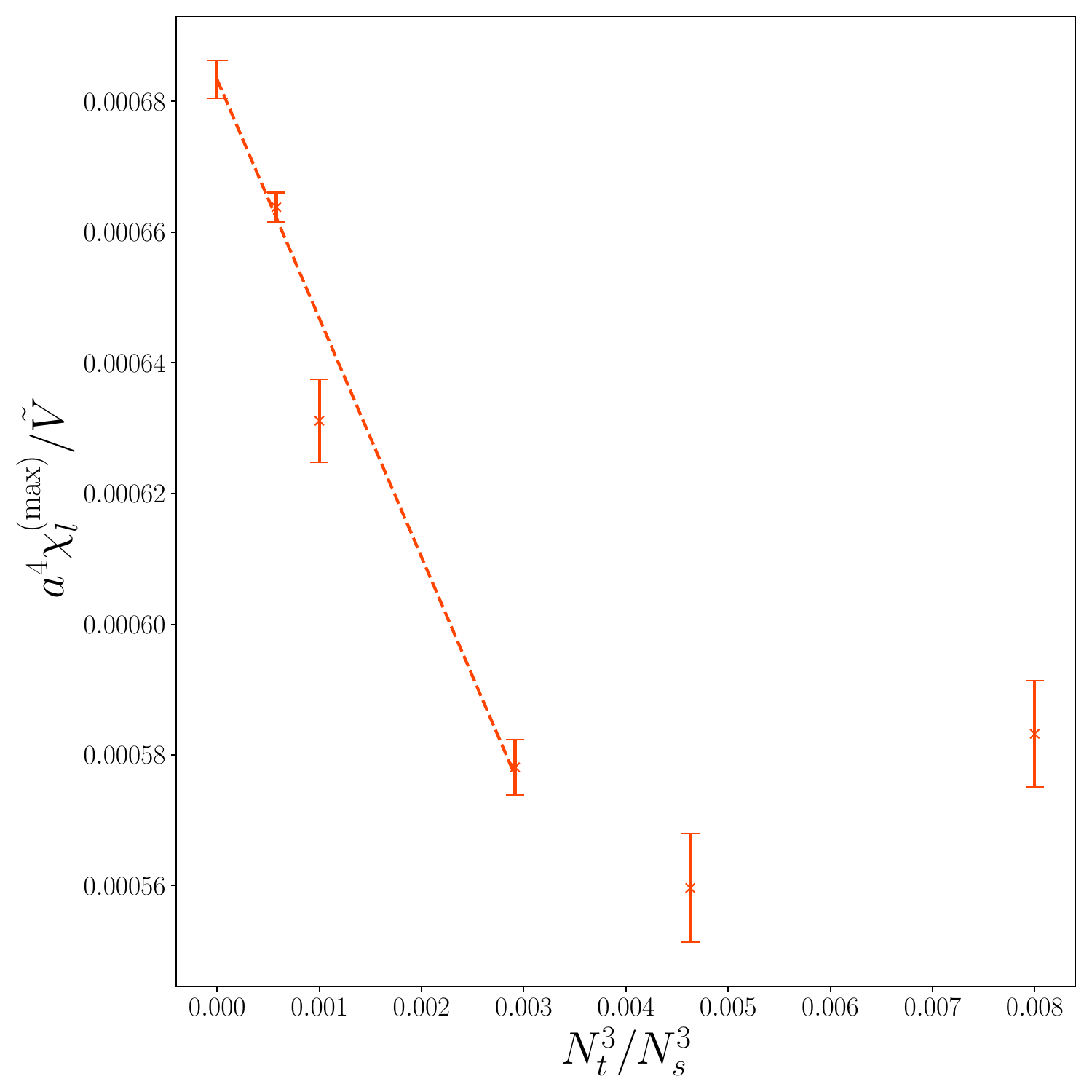}
\caption{\label{fig:XlMax_Nt4} Maxima of the Polyakov loop susceptibility, $\chi_l^\mathrm{(max)}$, as a function of the cube of the inverse of the aspect ratio, $N_t^3 / N_s^3$. The lattices considered have fixed temporal size, $N_t$ = 4, and spatial sizes $N_s = 20,\, 24,\, 28,\, 40,\, 48$.  For each spatial volume we use the extrapolations to the limit of vanishing subinterval size---see Appendix~\ref{App:DELim}. The black dashed curve shows a  quadratic polynomial fit to the results for the three largest spatial volumes. The fit had reduced chi-square $\chi^2/N_{\rm d.o.f.} = 6.7$. The thermodynamic limit, $N_s \to \infty$, yields $a^4 \chi_l^\mathrm{(max)} / \tilde V = 6.83(3)(5) \times 10 ^{-4}$, and is displayed at the far left of the plot.}
\end{figure}

\begin{figure}[t!]
\centering
\includegraphics[width=0.45\textwidth]{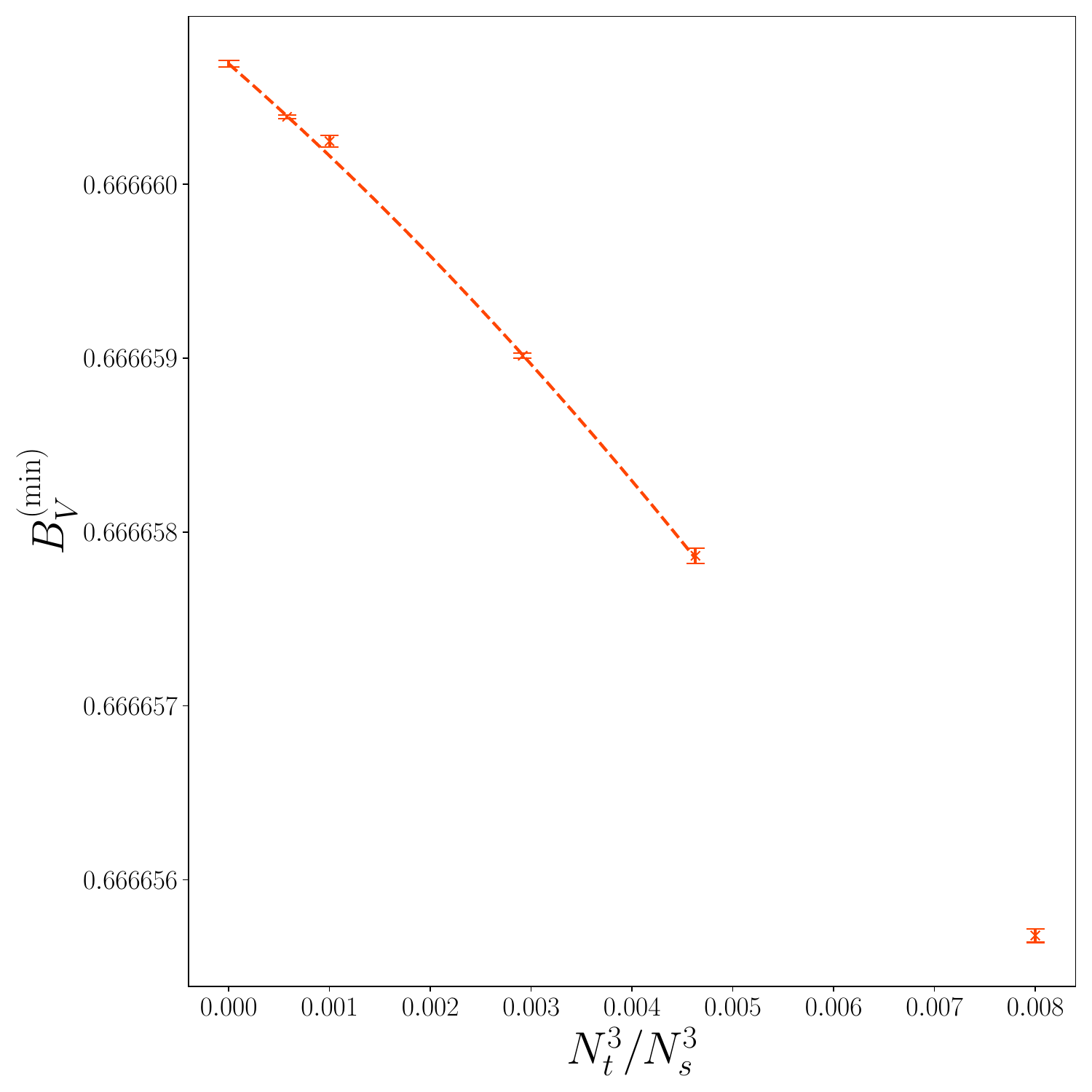}
\caption{\label{fig:BvMax_Nt4} Minima of the Binder cumulant, $B_V^\mathrm{(min)}$, as a function of the cube of the inverse of the aspect ratio, $N_t^3 / N_s^3$. The lattices considered have fixed temporal size, $N_t$ = 4, and spatial sizes $N_s = 20,\, 24,\, 28,\, 40,\, 48$.  For each spatial volume we use the extrapolations to the limit of vanishing subinterval size---see Appendix~\ref{App:DELim}. The black dashed curve shows a  quadratic polynomial fit to the results for the four largest spatial volumes.  The fit had reduced chi-square $\chi^2/N_{\rm d.o.f.}  = 6.2$. The thermodynamic limit, $N_s \to \infty$, yields 0.66666069(2)(1), and is displayed at the far left of the plot.}
\end{figure}

We have analysed our data by exploiting
 four alternative definitions for the critical point of the phase transition: the coupling for which the free energy of the two metastable branches are equal, $\beta_{CV}(f)$, the coupling corresponding to the peak of the specific heat, $\beta_{CV}(C_V)$, to the maximum of the Polyakov loop susceptibility,  $\beta_{CV}(\chi_l)$, and to the minimum of the Binder cumulant,  $\beta_{CV}(B_V)$.
Due to the presence of  finite volume effects, we expect discrepancies 
of order $\mathcal{O}(1/N_s^3)$ to appear at finite volume, but 
 we also expect all definitions to converge to one result in the thermodynamic limit. Nevertheless,
 while correlations must be present (as we measure the density of states just once, on a common
 set of numerical data),
 we can use the resulting four different estimates to assess the size of systematic effects 
 appearing due to the methodology adopted in the 
 measurement of the critical coupling. As we shall see, we find these systematic effects
 to have comparable size to the 
 statistical errors, and hence they cannot be neglected.

Figure~\ref{fig:Bc_Nslim_Nt4} shows the measurements  of these four critical couplings, for each available spatial volume, as a function of the cube of the inverse of the aspect ratio, $N_t^3/ N_s^3 $. In each case, we have first extrapolated towards the limit of vanishing interval size---see Appendix~\ref{App:DELim}. For each definition, we fit the measurements to a linear function (excluding the measurement with $N_s = 20$). We extrapolate all fits to the limit $N_t/ N_s \to 0$.  The extrapolation for the peak of the specific heat and minima of the Binder cumulant yield $\beta_{C}(C_V) =7.340018(7)(5)$ and $\beta_{C}(B_V) = 7.340019(7)(5)$,  with  reduced chi-squared of $\chi^2/N_{\rm d.o.f.} = 0.84$ and $\chi^2/N_{\rm d.o.f.} = 0.81$ respectively. The resulting agreement is not surprising, as both definitions are based on the behaviour of the average plaquette; 
even at finite volume the measurements are consistent with each other. 

The extrapolation of the critical coupling taken from the peak of the Polyakov loop susceptibility yields $\beta_{C}(\chi_l) = 7.340052(8)(4)$, with  reduced chi-square of $\chi^2/N_{\rm d.o.f.} = 0.26$. This lies  outside the error of the specific heat and Binder cumulant extrapolation, however the final point, at finite volume, is within error of the others. This discrepancy could be due the extrapolations not considering the higher-order effects on the finite-size scaling, due to the presence of mixed phase states.

The final extrapolation to the thermodynamic limit is given by $\beta_{C}(f) =7.340089(8)(2)$, which has a reduced chi-square of $\chi^2/N_{\rm d.o.f} = 0.08$. The reduced chi-square is very small in this fit and might be due to an overestimation of the errors. The result for this definition of the critical coupling is not consistent with the other ones, at either finite volume nor after extrapolations.

Another expectation of the thermodynamic limit, in the presence of   first-order phase transitions, and  based on the double Gaussian approximation, is that the peak of the specific heat, $C_V^\mathrm{(max)}$, should approach a value proportional to the square of the plaquette jump---see  Eq.~(\ref{eq:spec_to_latheat}). In Fig.~\ref{fig:DE2_Nslim_Nt4}, we compare the square of the measured value of the plaquette jump, found directly from the plaquette distribution at the critical point, with the peak of the specific heat, for all spatial sizes. In both cases we take the limit of vanishing interval size, as discussed in Appendix \ref{App:DELim}.

To test the theoretical expectation, we fit the specific heat term, $4a^4 C_V^\mathrm{(max)} / (6 \tilde V)$, to a quadratic polynomial in the cube of the inverse of the aspect ratio, $N_t^3 / N_s^3$. We then extrapolate the fitting result to the limit $N_t / N_s \to 0$. The reduced chi-square of the fit is $\chi/N_{\rm d.o.f.}^2 = 5.5$. The extrapolation is $4a^4 C_V^\mathrm{(max)} / (6 \tilde V) = 5.85(2)(1)\times 10 ^{-6}$. We also fit a function of the same form to the square of the plaquette discontinuity, $\Delta \langle u_p \rangle_{\beta_{CV}}^2$. The reduced chi-square is, in this case, $\chi^2/N_{\rm d.o.f.} = 0.18$ and the extrapolation is $\Delta \langle u_p \rangle_{\beta_C}^2 = 6.09(7)\times 10 ^{-6}$. These extrapolations don't agree, yet for the two largest volumes, $N_s = 40$ and $48$, both observables are consistent. This could indicate that higher order effects of the scaling need to be considered to improve the fit process itself. Further measurements at additional values of $N_s$ may be required, to verify if a divergence between the observables grows, or if they do tend to the same value.

We conclude with two final measurements. 
First, the Polyakov loop susceptibility evaluated at criticality is expected to scale linearly with the volume. In Fig.~\ref{fig:XlMax_Nt4}, we plot the value of the peak of the Polyakov loop susceptibility normalised by the volume, $a^4 \chi_l^\mathrm{(max)} / \tilde V$, against the cube of the inverse of the aspect ratio. We fit a linear function in $N_t^3/N_s^3$ to the results for $N_s \geq 28$. We find a  relatively large reduced chi-square $\chi^2/N_{\rm d.o.f} = 6.7$. The extrapolation to the thermodynamic limit is given by $a^4 \chi_l^\mathrm{(max)} / \tilde V = 6.83(3)(5) \times 10 ^{-4}$.

In Fig.~\ref{fig:Bv_Nt4}, we plot the value of the minimum of the Binder cumulant, $B_V^\mathrm{(min)}$, as a function of the cube of the inverse of the aspect ratio. We fit the measurements to a quadratic function in the cube of the inverse of the aspect ratio. The reduced chi-square for this fit is $\chi^2/N_{\rm d.o.f.} = 6.2$. The extrapolation to the limit $N_t / N_s \to 0$ is given by  $B_V^\mathrm{(min)}=0.66666069(2)(1)$. The value of $2/3$ (expected in second-order phase transitions) lies outside the error of this measurement. The reduced chi-square for this fit is large, as it might be affected by sub-leading corrections to the double-Gaussian approximation which arise from the mixed phase and the interface contributions.

While the statistical error on each or our measurements demonstrates the potential of the LLR algorithm to improve by several order of magnitude the precision of the measurements in the literature, however sizable systematic errors are present on the extrapolations to the thermodynamic limit. This can be seen by the comparably high reduced chi-square for the fits of the maximum of the specific heat, $C_V^\mathrm{(max)}$, the minima of the Binder cumulant, $B_V^\mathrm{(min)}$ and the maxima of the Polyakov loop susceptibility, $\chi_l^\mathrm{(max)}$. It can also be seen from the disagreement between different extrapolations, for example in the relation between $C_V^\mathrm{(max)}$ and the plaquette jump, and the different definitions of the critical couplings $\beta_C$. One possible explanation for these effects is that the double Gaussian approximation might be not accurate enough to describe the scaling at larger volumes, as it ignores mixed phase configurations on the observables.

\section{Outlook}
\label{Sec:outlook}

We applied the LLR algorithm to perform our first extensive, high-statistics  numerical study of the finite-temperature, confinement/deconfinement phase transition in the $Sp(4)$ lattice gauge theory.
For each of the lattice volumes available, we measured the three main quantities that characterise the transition: the critical coupling, $\beta_C$, the discontinuity in the plaquette, $\Delta \langle u_p \rangle_{\beta_C}$, and the quantity, $\tilde{I}$, that in the thermodynamic limit yields the interface tension, $\lim_{\frac{N_t}{N_s}\rightarrow 0} \tilde{I} = \tilde{\sigma}_{cd}$.

The first step of our analysis was to apply the LLR algorithm to reconstruct the density of states of the theory. Having done so, we measured  the free energy density, $f$. We demonstrated its multivalued nature, as expected in a first-order phase transition, in the region near criticality. 
We identified two competing metastable phases and one unstable mixed phase. 
We proceeded to compute other phenomenological observables, such as the probability distribution of the plaquette, the VEV of the average plaquette, the interphase contribution to the dynamics, the VEV of the
absolute value of the Polyakov loop, as well as its susceptibility, and the Binder cumulant.

We performed our numerical studies on a number of lattices with different volumes. For each, we rerun the LLR algorithm with several different choices of the extent of the subinterval, $\Delta u_p$, used in defining the log-linear approximation, $\tilde{\rho}$, to the true density of states of the theory, $\rho$. 
We then performed two extrapolations. For each lattice volume, we combined all the measurements obtained with different 
extent of the subintervals, $\Delta_{u_p}$, and extrapolated to the limit $\Delta u_p\rightarrow 0$, so that $\tilde{\rho}\rightarrow \rho$.
We then considered  the thermodynamic limit: holding fixed the temporal extent of the lattice, $N_t=4$, we extrapolated all the measurements to the infinite volume limit in spatial extent, $N_s\rightarrow +\infty$.

We obtained four determinations of the critical coupling in the $Sp(4)$  lattice gauge theory, for $N_t=4$, 
 in $1+3$ dimensions, which we repeat here:
\beqs
\beta_C(f)&=&7.340089(8)(2)\nonumber\,,\\
\beta_C(C_V)&=&7.340018(7)(5)\nonumber\,,\\
\beta_C(B_V)&=&7.340019(7)(5)\nonumber\,,\\
\beta_C(\chi_l)&=&7.340052(8)(4)\nonumber\,.
\eeqs
The labels of the four determinations 
refer to the  different observables used in the extraction of the critical coupling.
Our results are compatible with early measurement found in the literature,
 $\beta_C=7.339(1)$~\cite{Holland:2003kg}.
 But the statistical error, which is obtained by repeating the LLR measurement with $n_R$ replicas, and then bootstrapping the results, is two orders of magnitudes smaller than that quoted in the literature, which demonstrates the potential of the LLR method as a  precision tool.
 We also measured, for $N_t=4$, the infinite-volume extrapolations
 \beqs
 \lim_{N_t/N_s\rightarrow 0} \Delta \langle u_p \rangle_{\beta_C} &=& 0.002468(13)(1)\,,\nonumber\\
  \lim_{N_t/N_s\rightarrow 0} \tilde{I} &=& 0.01718(32)(1)\,.\nonumber
  \eeqs
These two quantities, discussed in the body of the paper, are related to latent heat and interface tension.

However,  the flexibility and precision provided by the LLR algorithm also ended up exposing the presence of unknown systematics in the infinite-volume extrapolations. In particular, the four determinations of $\beta_C$ are not compatible with one another, within statistical errors.  The systematic uncertainty as estimated in this work (see Appendices ~\ref{App:DELim} and ~\ref{App:Systematic}) only partially addresses the discrepancy. Resolving the latter is then the first target of future high precision studies. We envision to use a combination of large volumes to gain better control on the approach to the thermodynamic limit,  in the four observables of interest.

The next step requires to remove discretisation effects, and  extrapolate to the continuum limit. To this purpose, we envision repeating the same analysis, with different choices of $N_t$, and combine the results. At the time of writing this paper, we have started exploratory runs for $N_t=5$---see some preliminary results in Appendix~\ref{App:Nt5}.
This programme is going to be computationally challenging, as resolving the critical region near the transition for large choices of $N_t$ requires performing the whole LLR study with increasingly large spatial volumes, $N_s^3$.

Our third goal for future investigation is quite ambitious. This whole analysis exposed here
 can be repeated for other theories. Our collaboration is particularly interested in  the $SU(3)$ Yang-Mills theory~\cite{Mason:2022trc,Mason:2022aka,Lucini:2023irm}, but other gauge groups are of interest as well~\cite{Springer:2021liy,
   Springer:2022qos,Springer:2023wok,Springer:2023hcc},
and it would be equally interesting to consider theories coupled to fermions, besides the Yang-Mills dynamics.
These are challenging calculations, but the present study demonstrates that the LLR algorithm has the potential to lead to a revolutionary improvement in our control of the physics near the transition, which might have both theoretical and phenomenological implications, in particle as in astroparticle physics.

\acknowledgments
We would like to thank F.~Zierler for useful discussions on future developments of the software used for this work. We also thank Deog Ki Hong, Jong-Wan Lee and Fabian Zierler for comments on early versions of the manuscript.

The work of E.~B. has been funded by the UKRI Science and Technologies Facilities Council (STFC) Research Software Engineering Fellowship EP/V052489/1, by the STFC under Consolidated Grant No. ST/X000648/1, and by the ExaTEPP project EP/X017168/1. 
The work of D.~V. is supported by STFC under Consolidated Grant No.~ST/X000680/1.
The work of D.~M. is supported by a studentship awarded by the Data Intensive Centre for Doctoral Training, which is funded by the STFC grant ST/P006779/1.  
B.~L. and M.~P. received funding from the European Research Council (ERC) under the European Union’s Horizon 2020 research and innovation program under Grant Agreement No.~813942, and by STFC under Consolidated Grants No. ST/P00055X/1, ST/T000813/1, and ST/X000648/1. 
The work of B.~L. is further supported in part by the Royal Society Wolfson Research Merit Award WM170010 and by the Leverhulme Trust Research Fellowship No. RF-2020-4619. 

Numerical simulations have been performed on the Swansea SUNBIRD cluster (part of the Supercomputing Wales project) and AccelerateAI A100 GPU system.The Swansea SUNBIRD system and AccelerateAI are part funded by the European Regional Development Fund (ERDF) via Welsh Government. 

Numerical simulations have been  performed 
on the DiRAC Data Intensive service at Leicester. The DiRAC Data Intensive service equipment at Leicester was funded by BEIS capital funding via STFC capital grants ST/K000373/1 and ST/R002363/1 and STFC DiRAC Operations grant ST/R001014/1. 

Numerical simulations have used the DiRAC Extreme Scaling service at the University of Edinburgh. The DiRAC Data Intensive service at Leicester is operated by the University of Leicester IT Services, and the DiRAC Extreme Scaling service is operated by the Edinburgh Parallel Computing Centre, they form part of the STFC DiRAC HPC Facility (www.dirac.ac.uk). 

This work used the DiRAC Data Intensive service (CSD3) at the University of Cambridge, managed by the University of Cambridge University Information Services on behalf of the STFC DiRAC HPC Facility (www.dirac.ac.uk). The DiRAC component of CSD3 at Cambridge was funded by BEIS, UKRI and STFC capital funding and STFC operations grants. DiRAC is part of the UKRI Digital Research Infrastructure.
The DiRAC Extreme Scaling service was funded by BEIS capital funding via STFC capital grant ST/R00238X/1 and STFC DiRAC Operations grant ST/R001006/1.
DiRAC is part of the National e-Infrastructure.

This work was supported by the Supercomputer Fugaku Start-up Utilization Program of RIKEN.
This work used computational resources of the supercomputer Fugaku provided by RIKEN through the HPCI System Research Project (Project ID: hp230397).

{\bf Open Access Statement - } For the purpose of open access, the authors have applied a Creative Commons 
Attribution (CC BY) licence  to any Author Accepted Manuscript version arising.

{\bf Research Data Access Statement}---The data and analysis code for this manuscript can be downloaded from  Ref.~\cite{DataRelease1}. The simulation code can be found from Ref.~\cite{DataRelease2}.

\appendix
\section{LLR algorithm}
\label{App:LLR}
The coefficients, $a_n$, defined in Eq.~(\ref{eq:piecewise}), are found by solving numerically, 
for each $a_n$, the equation
\beqs
0&=&
\langle \langle \Delta E \rangle \rangle_ n (a_n) \nonumber\\
&=&  \frac{1}{{\cal N}_n(a_n)} \int_{E_n-\frac{\Delta_E}{2}}^{E_n+\frac{\Delta_E}{2}} dE (E - E_n) {\rho}(E)
e^{-a E} ,
\eeqs
in each interval, $n=1,\,\cdots,\,N_{\rm int}$,  with
\beq
{\cal N}_n(a_n) \equiv \int_{E_n-\frac{\Delta_E}{2}}^{E_n+\frac{\Delta_E}{2}} {\rho}(E)
e^{-a_n E} dE \,.
\eeq
The double angle brackets denote the expectation value of an observable computed with energy restricted to an interval, $E_n -\Delta_E/2 < E < E_n + \Delta_E/2$. This interval differs from the interval in Eq.~(\ref{eq:piecewise}), to allow for an overlap region between adjacent intervals for replica exchange as we will discus shortly.

We solve this equation iteratively, by using a combination of Newton-Raphson (NR) and  Robbins-Monro (RM) methods. We denote our approximations as $a_n^{(m)} \approx a_n$, where $(m)$ is the iteration number. Each NR or RM update then takes this estimate and improves it, taking $a_n^{(m)} \to a_n^{(m+1)}$, having started from an initial trial value, $a_n^{(0)}$. The NR iteration is given by
\beq
\label{eq:nr}
a_n^{(m+1)} = a_n^{(m)} - \frac{12}{\Delta_E^2} \langle \langle \Delta E \rangle \rangle_ n (a_n^{(m)}) \,,
\eeq
while the RM method increments are finer, with
\beq
\label{eq:rm}
a_n^{(m+1)} = a_n^{(m)} - \frac{12}{(m+1) \Delta_E^2} \langle \langle \Delta E \rangle \rangle_ n (a_n^{(m)})\,.
\eeq
The NR iterations bring the system close to the true value more quickly than the RM iteration, but the RM iteration are required to refine the accuracy of the estimate.

The double-angle-bracket expectation values are recalculated at each step, using configurations generated by a combination of over-relaxation and heat-bath steps, with energy restricted to the prescribed interval. 
The restricted heat-bath algorithm is discussed in Appendix~A of Ref.~\cite{Lucini:2023irm}, which we adapted to symplectic gauge groups, as discussed in Appendix~\ref{App:HB_Append}. For each NR and RM iteration, we generate $n_{\rm Th}$ configurations to reduce autocorrelation, then measure $\langle \langle \Delta E \rangle \rangle_ n (a)$ using $n_M$ configurations. For lattices with $N_t=4$, the values of the parameters $n_{\rm Th}$ and  $n_M$ are listed in Table~\ref{tab:lat_surf1} .

The lattice is decomposed into $N_D$ subdomains. To account for the fact that the ensembles we produce are defined by a the global constraint on the energy, we perform a restricted heat-bath update on one subdomain at a time, chosen on a rota, while on the other subdomains we carry out an over-relaxation update. One full lattice update is complete when each subdomain has undergone one restricted heat-bath update.

To ensure ergodicity, we determine $a_n$ in all intervals simultaneously, by running one independent lattice for each interval. Between each (NR or RM) iteration, we introduce the possibility of a swap between two lattices in adjacent intervals, if they are both in the shared overlap region. The swaps are stochastically determined, with probability
\begin{eqnarray}
\label{eq:swap}
  P_{\mathrm{swap}} = \min \left( 1, e^{(a_n^{(m)} -
  a_{n-1}^{(m)})(E_n^{(m)}- E_{n-1}^{(m)} )} \right) \,.
\end{eqnarray}
Furthermore, we smoothen the sharp cutoffs in the distribution appearing at the end points of the total interval. We do so in the first (last) interval $E_1 -\Delta_E/2 < E < E_1 + \Delta_E/2$ (and $E_{N_{\rm int}} -\Delta_E/2 < E < E_{N_{\rm int}} + \Delta_E/2$ ), by  replacing the boundaries with $0 < E < E_1 + \Delta_E/2$ (and $E_{N_{\rm int}} -\Delta_E/2 < E < 6 \tilde{V}/a^4$). In this case the NR and RM iterations cannot be used. Instead, we assume that at the boundary $a_n$ changes approximately linearly, making $a_1 \equiv 2a_2 - a_3$ and $a_{N_{\rm int}} \equiv 2a_{N_{\rm int}-1} - a_{N_{\rm int}-2}$. 

As explained in the body of the paper, starting points, $a_n^{(0)}$, are determined through a rough preliminary calculation using important sampling methods. A systematic error is introduced by the truncation of the iterative process after $\bar{m}$ NR and  $\tilde{m}$
 RM iterations. We estimate its size
by repeating our determination of $a_n$ with a different random seed, and measuring the distribution of the results. 
As each replica begins in a randomly generated configuration,  before we can use the action in constrained updates, the energy of the lattice system must be brought into the relevant interval. We refer to this process as annealing. This is achieved by using standard heat-bath updates with the coupling set to the initial value of $a_n$. After each update, we check if the energy is within the relevant interval, if it is we have finished annealing. After a set number of updates, if the system is still not within the interval, the coupling is changed---it is increased if $E<E_n$ and decreased if $E>E_n$.

Once the RM iterations have completed, our final values, $\{a_n^{(\tilde{m})}\approx a_n\}_{n=1}^{N_{\rm int}}$, have been determined for each interval centered at $\{E_n\}_{n=1}^{N_{\rm int}}$. We then carry out a set of restricted heat-bath updates. On each configuration, we measure the Polyakov loop and the average plaquette. To maintain ergodicity, after $n_\text{fxa}$ iterations we apply a replica exchange, and further  repeat the process $\hat m$ times. This yields $n_\text{fxa} \times \hat m$ measurements in each interval to estimate Eq.~(\ref{eqn:vev_gen_small}) for an arbitrary value of  $\beta$. We hence arrive at the expectation value in Eq.~(\ref{eqn:vev_gen}). The error on this VEV is determined by bootstrapping over the results from each replica.

We conclude by providing a schematic description of the whole algorithm.
\begin{itemize}
\item[A.] 
Choose the energy range physically relevant to the problem, $[E_{\mathrm{min}}, E_{\mathrm{max}}]$, and divide it into $N_{\rm int}$ subintervals, with energies centred at $E_n = E_{\mathrm{min}} + n \Delta_E/2$, for $n=1,..,N_{\rm int}$.
Define initial values, $a_n^{(0)}$, for each of the intervals.
\item[B.] Repeat $n_R$ times with different random sequences:
  \begin{enumerate}
  \item Start replica in a random configuration and anneal. Then thermalise with action-constrained updates.
  \item Repeat $\bar{m}$ times:
   \begin{enumerate}
   \item Thermalise with $n_{\rm Th}$ action-constrained updates.
   \item Measure $(E-E_n)$ on $n_M$ action-constrained updates to calculate the VEV in Eq.~(\ref{eq:nr}) and update $a_n$.
   \item Consider configuration swaps using Eq.~(\ref{eq:swap}).
   \end{enumerate}
     \item Repeat $\tilde{m}$ times:
   \begin{enumerate}
   \item Thermalise with $n_{\rm Th}$ action-constrained updates.
   \item Measure $E$ on $n_M$ action-constrained updates to calculate the VEV in Eq.~(\ref{eq:rm}) and update $a_n$.
   \item Consider configuration swaps using Eq.~(\ref{eq:swap}).
  \end{enumerate}
   \item Repeat $\hat{m}$ times:
   \begin{enumerate}
   \item Identify $a_n$ with the final values, $a_n^{(\tilde{m})}$, generate $n_{\text{fxa}}$ for configurations restricted to each interval and measure an observable $O[U]$ on them. In this work $O[U] = l_p[U]$. 
   \item Consider configuration swaps using Eq.~(\ref{eq:swap}).
   \end{enumerate}
  \end{enumerate}
\end{itemize}

\begin{table*}
\caption{List of $Sp(4)$ lattice ensembles with extension in the time direction  $N_t=5$. 
For each of them, we list the lattice extension, $N_t \times N_s^3$,
 the number of (overlapping) subintervals, $N_{\rm int}$, 
the number of replicas, $n_R$, the extent  of the interval in average plaquette, $\Delta_{u_p}$,  and the maximum and minimum values of the plaquette, $(u_p)_{\mathrm{min}}$ and $(u_p)_{\mathrm{max}}$.
We also report the number of iterations defined in Appendix~\ref{App:LLR}: $\bar{m}$, 
$\tilde{m}$, $\hat{m}$, $n_{\text{Th}}$, $n_{M}$, and $n_{\text{fxa}}$. }
 \label{tab:lat_surf2}
\begin{center}
\begin{tabular}{|c|c|c|c|c|c|c|c|c|c|c|c|}
\hline
$N_t \times N_s^3$ & $N_{\mathrm{int}}$ & $n_R$ & $\Delta_{u_p}$ & $(u_p)_{\mathrm{min}}$ & $(u_p)_{\mathrm{max}}$ & $\bar{m}$ & $\tilde{m}$ & $\hat{m}$ & $n_{\mathrm{Th}}$ & $n_M$ & $n_{\mathrm{fxa}}$ \\
\hline
$5 \times 48^3$ & 48 & 25 & 0.00017 & 0.5875 & 0.5915 & 10 & 60 & 50 & 300 & 700 & 100 \\
$5 \times 48^3$ & 96 & 25 & 0.00008 & 0.5875 & 0.5915 & 10 & 50 & 50 & 300 & 700 & 100 \\
$5 \times 56^3$ & 48 & 25 & 0.00017 & 0.5875 & 0.5915 & 10 & 50 & 50 & 300 & 700 & 100 \\
$5 \times 56^3$ & 96 & 25 & 0.00008 & 0.5875 & 0.5915 & 10 & 50 & 50 & 300 & 700 & 100 \\
$5 \times 56^3$ & 128 & 25 & 0.00006 & 0.5875 & 0.5915 & 10 & 50 & 50 & 300 & 700 & 100 \\
\hline
\end{tabular}

\end{center}
\end{table*}

\section{Restricted heat-bath algorithm for $Sp(2N)$ gauge groups}\label{App:HB_Append}
The link variables of the lattice are sampled following the prescription outlined in Ref.~\cite{Cabibbo:1982zn}, and extended to $Sp(2N)$ gauge groups in Ref.~\cite{Bennett:2020qtj}. The  $Sp(2N)$ elements are generated by updating $SU(2)\sim Sp(2)$ subgroups, themselves sampled through energy-restricted heat-bath updates, as discussed in Appendix A of Ref.~\cite{Lucini:2023irm}. 
Given a link variable $U\in SU(2)$, and $U_\sqcup$  the staple around $U$, which has determinant  $k$,
the updated link is then given by $U'=U U_\sqcup / k$.
In order to determine $U$, we use the standard $SU(2)$ parameterisation of the non-trivial part of the embedding as $U=u_0\mathbb{I}_2+i\,\vec{u}\cdot\vec\tau$, where $\vec\tau$ are the Pauli matrices and $\mathbb{I}_2$ is the 2x2 identity matrix.  
We sample uniformly $\vec{u}$ from a sphere of radius $\sqrt{1-u_0^2}$.
Finally,  $u_0$ is sampled by randomly generating a real variable, $0\leq \xi \leq 1$, and defining
\beq
\label{eq:a0distrib_1}
u_0 = \frac{1}{\beta k }
    \log{\left( e^{-\beta k u_\mathrm{min}} + \xi ( e^{\beta k u_\mathrm{max}}-e^{-\beta k u_\mathrm{min}}) \right)}\,,
\eeq
where $u_\mathrm{min}$ and $u_\mathrm{max}$ are determined imposing the constraints $E_{n-1}\leq E - E_i + E_f \leq E_{n+1}$, where $E$ is the total energy of the original configuration, $E_i$ is the contribution to $E$ given by the old link and $E_f$ is the energy contribution given by the new link.

The $SU(2)$ subgroups of $Sp(2N)$ can have two different structures. For instance, for $Sp(4)$ they can be either of the form
\beq 
\label{AppB:subgroup1}
~ 
\begin{pmatrix}

u_0 + iu_3 & 0 & u_1 - iu_2 & 0\\
0 & 1 & 0 & 0 \\
- u_1 - iu_2 & 0 & u_0 - iu_3 & 0 \\
0 & 0 & 0 & 1 \\
\end{pmatrix} 
\eeq
or of the form
\beq 
\label{AppB:subgroup2}
~ 
\begin{pmatrix}
u_0 + iu_3 & u_1 - iu_2 & 0 & 0\\
- u_1 - iu_2 & u_0 - iu_3 & 0 & 0 \\
0 & 0 & u_0 - iu_3 & u_1 + iu_2 \\
0 & 0 & - u_1 + iu_2 & u_0 + iu_3 \\
\end{pmatrix}
\ . 
\eeq
In both cases, we can write $E_f = 2(d-1) - kz u_0$, where $z$ will change depending on the trace of the specific subgroup\footnote{This formula also applies to $SU(N)$, for which $z = 1$.}. We then set $z = 1$ for the subgroups given in Eq.~(\ref{AppB:subgroup1}) and $z = 2$ for the subgroups given in Eq.~(\ref{AppB:subgroup2}). With these conventions, the constraints of $u_0$ become
\begin{align}\label{eq:xminmax}
 u_{\mathrm{min}}&=\max{ \left(\frac{2(d-1)+ (E-E_{n+1}) -E_i}{kz},-1\right)}~,\\ 
 u_{\mathrm{max}}&=\min{\left(\frac{2(d-1) + (E-E_{n-1}) -E_i}{kz},1\right)}~.
\end{align}
One heat-bath sweep requires the sequential update of all links of the lattice using this procedure.

\section{$N_t = 5$}
\label{App:Nt5}
While this paper focuses on the measurements in the $Sp(4)$ lattice gauge theory with lattices having fixed
temporal extent, $N_t=4$, yet, our long term objective is to extrapolate our measurements toward the continuum limit. We hence started
 exploratory runs to repeat our analysis for different values of $N_t$. In this Appendix we present preliminary results for $N_t =5$. Using standard importance sampling methods, we determined that a spatial size of $N_s = 40$ 
is the minimum one required to resolve the double peak structure in the plaquette distribution. We hence chose to perform our preliminary analysis  with the spatial sizes $N_s = 48$ and $N_s=56$.

We carried out an importance sampling analysis on $Sp(4)$ pure gauge theory with lattice size $5\times 48^3$. We considered lattice couplings $\beta = 7.480,\,7.490,\, 7.493,\, 7.495$, and $7.4$. For each update we carry out one heat-bath and four over-relaxation steps, and generated $10,000$ configurations, after having disregarded $5000$ thermalisation steps. For each ensemble we measured the absolute value of the Polyakov loop and average plaquette. From the rough analysis of our results,  we decided to fix the plaquette range of interest to be $(u_p)_\mathrm{min} = 0.5875<u_p<(u_p)_\mathrm{max} = 0.5915$, and used a polinomyal fit of the plaquete to fix our initial estimates for the coefficients, $a_n^{(0)}$.  Details on the LLR analysis are given in Tab.~\ref{tab:lat_surf2}.

\begin{figure}[t!]
\centering
\includegraphics[width=0.45\textwidth]{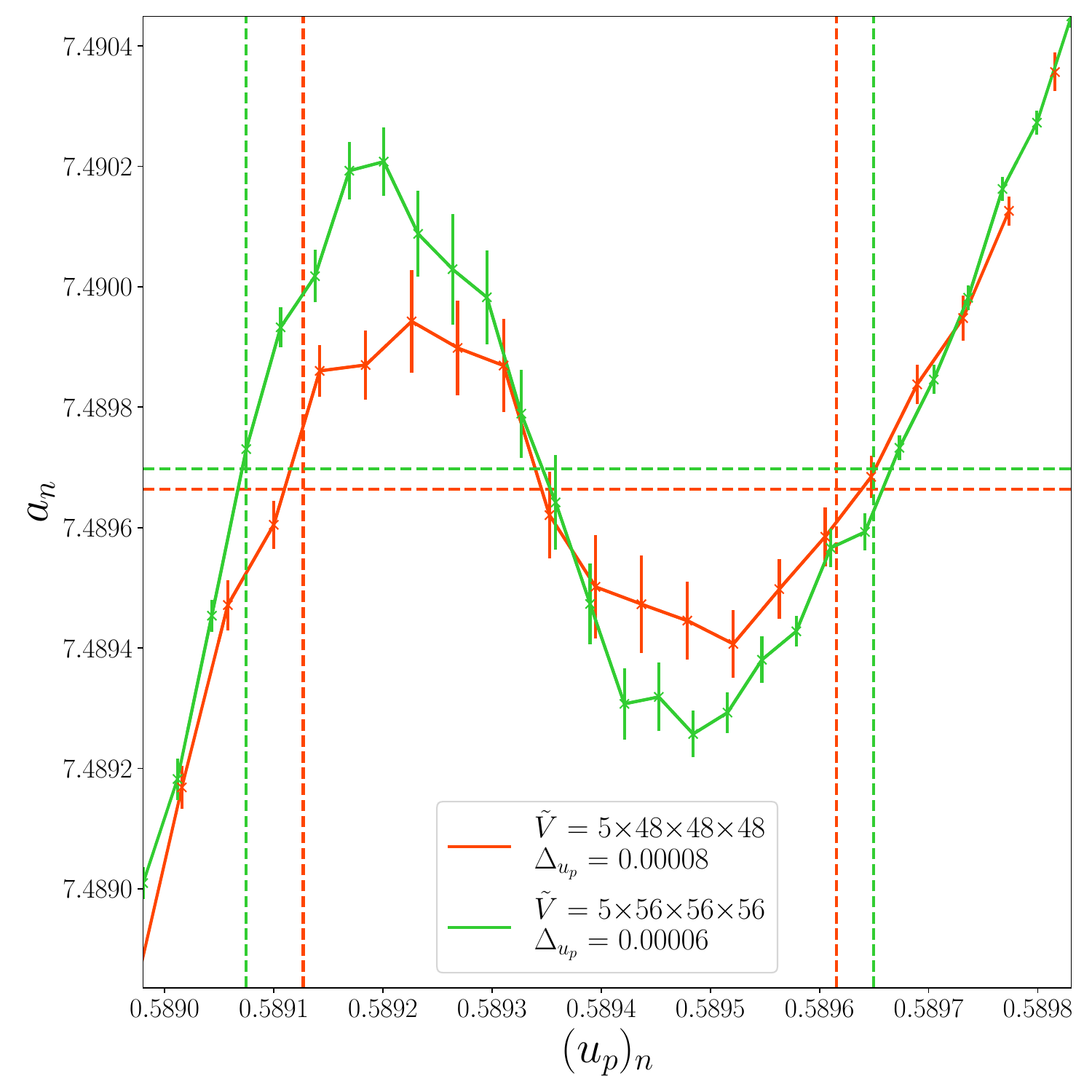}
\caption{\label{fig:an_En_Nt5}  The coefficients $a_n$, plotted as a function of the plaquette value at the centre of the energy interval, $(u_p)_n$, for the $Sp(4)$ theory on lattices with temporal size of $N_t=5$, spatial sizes $N_s = 48,\, 56$, and plaquette subinterval sizes $\Delta_{u_p} = 0.00008$, and $\Delta_{u_p} = 0.00006$, respectively, shown in different colours. The plot is focused on the critical region, showing the characteristic non-invertible structure of $a_n((u_p)_n)$. The horizontal dashed lines show the critical coupling, $a_n = \beta_{CV}(f)$, and the vertical lines show the corresponding plaquette values. }
\end{figure}

Our measurement of the coefficients, $a_n$, determining the density of states are plotted against the plaquette value at the centre of the interval, $(u_p)_n$, in Fig.~\ref{fig:an_En_Nt5}. The plot focuses on the critical region, in which we find clear evidence of  noninvertibility for the function $a_n(u_p)$.  
The horizontal dashed line shows the critical couplings relation to the coefficients, $a_n = \beta_{CV}(f)$, the vertical lines are the intercepts with the plaquette values, $u_p(a_n = \beta_{CV}(f))$ on the metastable branches. As for the case $N_t = 4$, we see that as the spatial volume grows, the range of $a_n$ values in the non-invertible region grows, indicating that the range of $\beta$ values that will be affected by metastable dynamics has grown.    
We notice that the value of the critical coupling is larger than for $N_t=4$.

\begin{figure}[t!]
\centering
\includegraphics[width=0.45\textwidth]{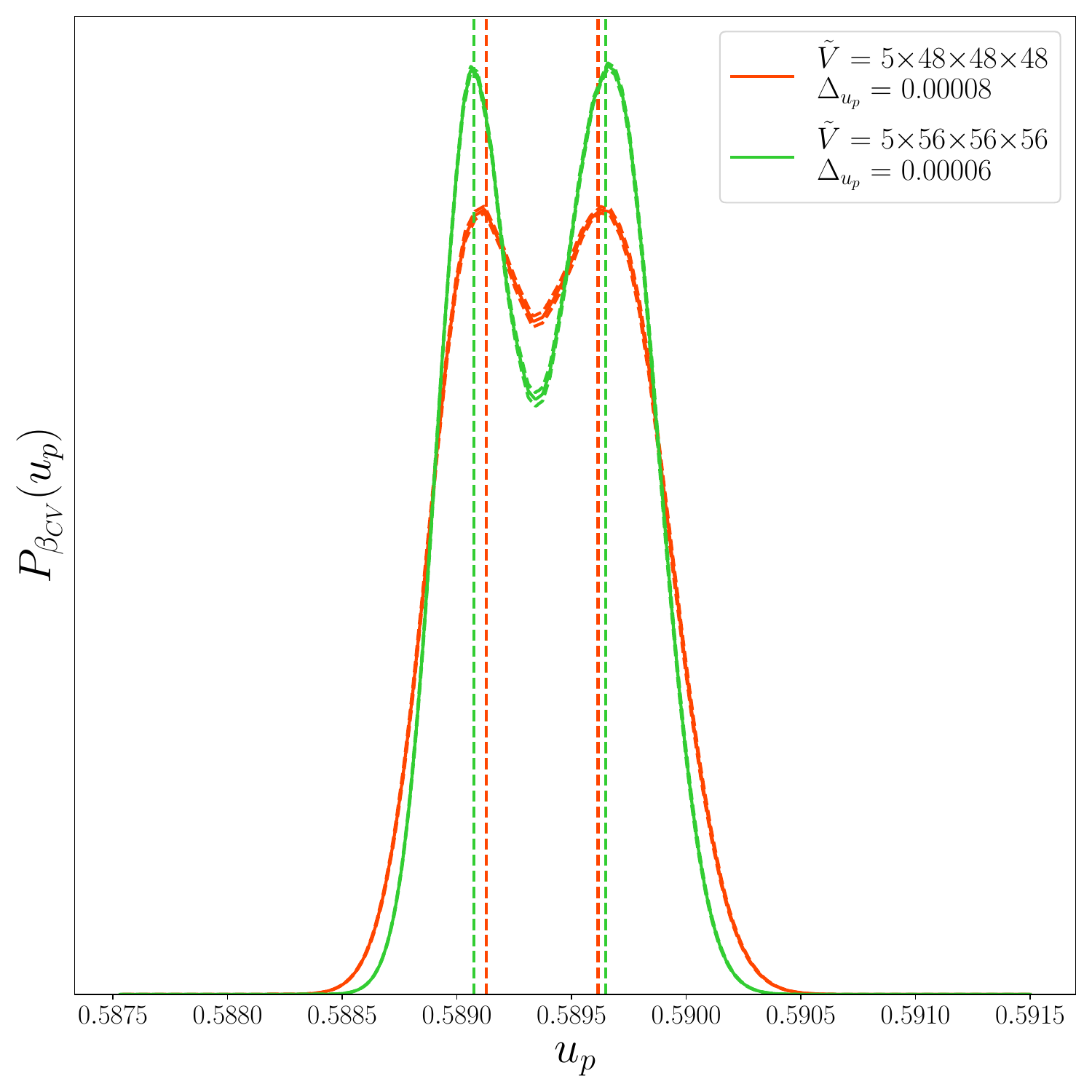}
\caption{\label{fig:PbE_Nt5} The plaquette probability distribution, $P_\beta(u_p)$, in the $Sp(4)$ lattice gauge theory with $N_t=5$, computed  at the critical point, $\beta=\beta_{CV}(f)$. The plaquette distribution exhibits a double Gaussian structure with peaks of equal height. 
As in Fig.~\ref{fig:an_En_Nt5} the color coding represents two different spatial sizes, $N_s= 48$ and $N_s=56$, with interval sizes $\Delta_{u_p} = 0.00008$ and   $\Delta_{u_p} =0.00006$ respectively. The horizontal dashed line shows the location of the peaks of the distribution. As the spatial volume is increases, the peaks become sharper.}
\end{figure}

\begin{figure}[t!]
\centering
\includegraphics[width=0.45\textwidth]{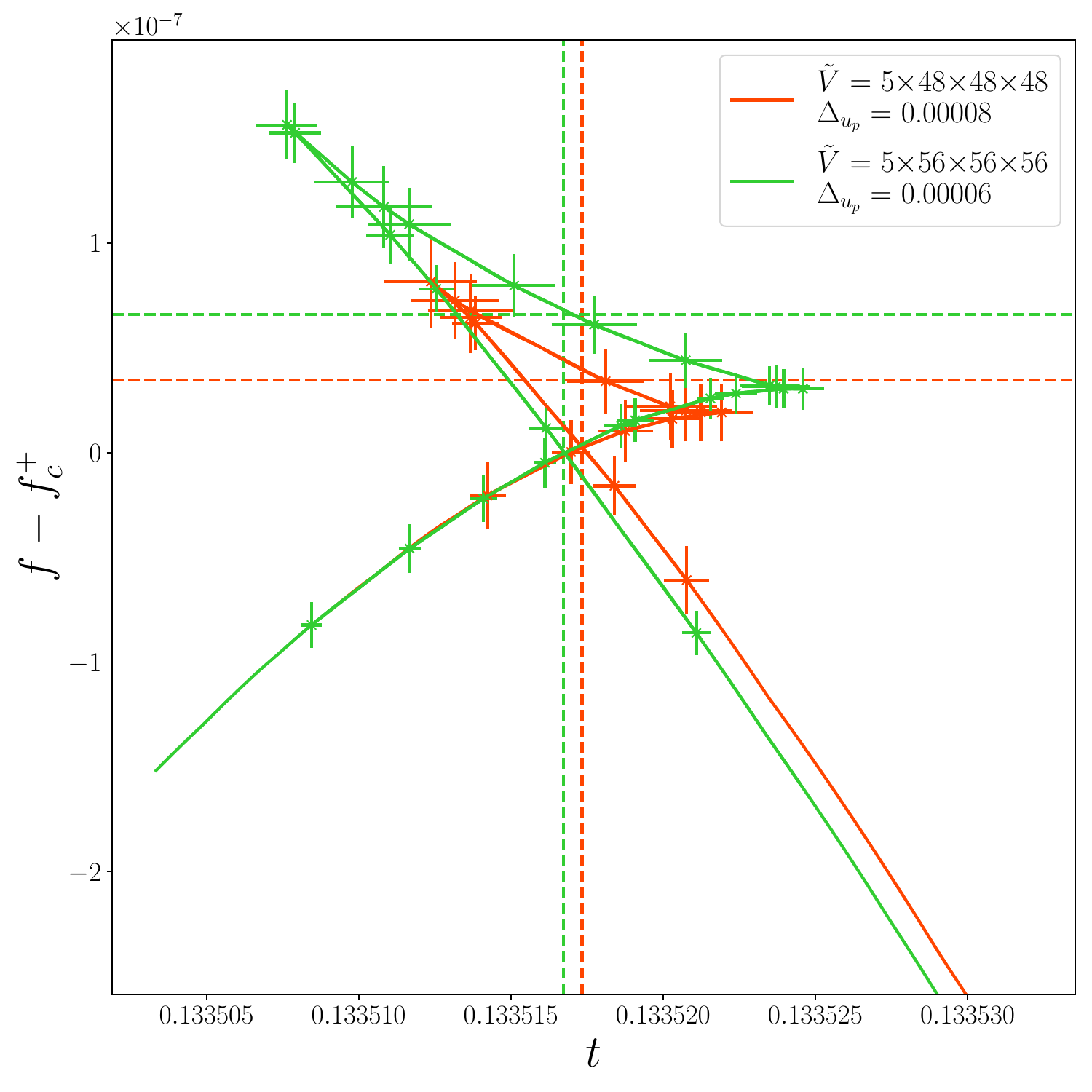}
\caption{\label{fig:t_F_Nt5} The reduced free energy, $f$, a function of the microcanonical temperature, $t=1/a_n$, for the $Sp(4)$ theory on lattices with temporal size $N_t = 5$, spatial sizes $N_s= 48$, and $N_s= 56$, and plaquette subinterval sizes $\Delta_{u_p} = 0.00008$, $\Delta_{u_p} = 0.00006$, respectively, as in Figs.~\ref{fig:an_En_Nt5} and~\ref{fig:PbE_Nt5}. To ensure that the (reduced) free energy difference can be  compared between different lattice sizes, constant term, $f_c^+$, has been subtracted, determined by the reduced free energy at which the two metastable branches cross.}
\end{figure}

But we also highlight that despite the large spatial value, the signal we detect is less pronounced. This is particularly clear when comparing the corresponding plaquette distributions, measured at the critical coupling, which are shown in Fig.~\ref{fig:PbE_Nt5}. As we increase the spatial volume, the peaks become more pronounced, yet for these choices of volume  the double Gaussian approximation is still accurate. We also show the reduced free-energy, $f(t)$, in Fig.~\ref{fig:t_F_Nt5}, in which we subtracted the value at the point at which the two metastable branches cross, as for the $N_t=4$ case.  We conclude that the choices of spatial volume applied to this preliminary study are  not close to the asymptotic regime, yet. Nevertheless, these measurements are sufficient to establish the first order nature of the transition, through both the double peak structure of the plaquette distribution and the swallow-tail structure of the free-energy.

In future work, we aim to extrapolate our analysis toward the continuum limit, which 
might be feasible if we are able to perform  the thermodynamic limit on lattices with both $N_t = 5$ and $N_t=6$. These are challenging goals, in view of the preliminary results shown here.
In the $N_t = 4$, we only required an aspect ratio of around $N_s / N_t = 5$, to resolve the separate peaks at the critical point of the first order phase transition. For $N_t = 5$, we required an aspect ratio  $N_s / N_t \sim 9.6$. An analysis of $N_t = 6$, therefore may require further increases of the aspect ratio. Further software development on the parallelisation of the algorithm may be required as well, before a satisfactory exploration of $N_t = 6$ case
is feasible. 

\section{$\Delta_E \to 0$ limit}
\label{App:DELim}
Measurements obtained with the LLR method are affected by  a systematic 
uncertainty related to the finite size of the subinterval in plaquette (or energy) of the system. One expects this effect to scale with the square of the interval size, $\Delta_E^2$~\cite{Langfeld:2015fua}. In this Appendix, we provide some additional details about the process we applied to accurately account for this source of uncertainty. For each of our measurements, we repeated the LLR process for multiple choice of subinterval sizes,
 and took the limit $\Delta_E \to 0$ of the results. 

\begin{table*}[t!]
\caption{ \label{tab:Nt4_critobs}  
Measurements in the $Sp(4)$ pure gauge theory on lattices with temporal size $N_t = 4$, performed at the critical coupling,
using the LLR algorithm, with subinterval size $\Delta_{u_p}$. Extrapolations to the limit of vanishing subinterval size are denoted by $\Delta_{u_p} \to 0$. The final extrapolations to the thermodynamic limit are denoted by $N_s \to \infty$. 
We denote  as
$a ^4 C_V^\mathrm{(max)} / \tilde V$ and $a^4 \chi_l^\mathrm{(max)}  / \tilde V$ 
the extrema of the specific heat, $C_V$, and the the Polyakov loop susceptibility, $\chi_l$, scaled by the lattice volume $\tilde V / a^4$, respectively.  The minima of the Binder cumulant are denoted as $B_V^\mathrm{(min)}$. The plaquette discontinuity, $\Delta \langle u_p \rangle_{\beta_{CV}}$ is measured as the difference between the two peaks of the plaquette distribution, when the peaks are of equal height. The quantity $\hat I$ is defined in Eq.~(\ref{eq:Interface}), and is related to the surface tension. For thermodynamic limit  we include an additional bracket containing an estimation for the systematic error. 
}
\begin{center}
\begin{tabular}{|c|c|c|c|c|c|c|}
\hline
$N_s$ &  $\Delta_{u_p}$ & $a ^4 C_V^\mathrm{(max)} / \tilde V$ & $B_V^\mathrm{(min)}$ & $a^4 \chi_l^\mathrm{(max)}  / \tilde V$ & $\Delta \langle u_p \rangle_{\beta_{CV}}$ & $\hat I$  \\
\hline
20 & 0.00064 & 1.611(3)$\times 10 ^{-5}$ & 0.66665571(2) & 5.86(4)$\times 10 ^{-4}$ & 0.002056(33) &  0.03135(10)\\
20 & 0.00048 & 1.614(2)$\times 10 ^{-5}$ & 0.66665570(1) & 5.85(3)$\times 10 ^{-4}$ & 0.002013(45) &  0.03148(8)\\
20 & $\to 0$ & 1.617(6)$\times 10 ^{-5}$ & 0.66665568(4) & 5.83(8)$\times 10 ^{-4}$ & 0.001958(109) &  0.03165(22)\\
24 & 0.00064 & 1.280(2)$\times 10 ^{-5}$ & 0.66665796(2) & 5.79(4)$\times 10 ^{-4}$ & 0.002352(10) &  0.02561(10)\\
24 & 0.00048 & 1.286(3)$\times 10 ^{-5}$ & 0.66665792(2) & 5.70(3)$\times 10 ^{-4}$ & 0.002375(12) &  0.02586(10)\\
24 & $\to 0$ & 1.294(7)$\times 10 ^{-5}$ & 0.66665786(4) & 5.60(8)$\times 10 ^{-4}$ & 0.002403(30) &  0.02616(25)\\
28 & 0.00048 & 1.109(2)$\times 10 ^{-5}$ & 0.66665912(2) & 5.69(4)$\times 10 ^{-4}$ & 0.002461(8) &  0.02278(8)\\
28 & 0.00025 & 1.120(1)$\times 10 ^{-5}$ & 0.66665905(1) & 5.76(3)$\times 10 ^{-4}$ & 0.002472(10) &  0.02318(5)\\
28 & $\to 0$ & 1.125(2)$\times 10 ^{-5}$ & 0.66665901(1) & 5.78(4)$\times 10 ^{-4}$ & 0.002477(14) &  0.02334(8)\\
40 & 0.00017 & 0.944(2)$\times 10 ^{-5}$ & 0.66666024(1) & 6.21(2)$\times 10 ^{-4}$ & 0.002509(6) &  0.01973(8)\\
40 & 0.00013 & 0.944(2)$\times 10 ^{-5}$ & 0.66666024(1) & 6.25(2)$\times 10 ^{-4}$ & 0.002497(7) &  0.01969(8)\\
40 & $\to 0$ & 0.943(5)$\times 10 ^{-5}$ & 0.66666025(3) & 6.31(6)$\times 10 ^{-4}$ & 0.002481(18) &  0.01965(20)\\
48 & 0.00013 & 0.922(1)$\times 10 ^{-5}$ & 0.66666039(1) & 6.64(2)$\times 10 ^{-4}$ & 0.002482(5) &  0.01897(8)\\
$\to \infty$&& 0.878(3)(1)$\times 10 ^{-5}$ & 0.66666069(2)(1) & 6.83(3)(5)$\times 10 ^{-4}$ & 0.002468(13)(1) &  0.01718(32)(1)\\
\hline
\end{tabular}

\end{center}
\end{table*}

We list in Tab.~\ref{tab:Nt4_critobs} and Tab.~\ref{tab:Nt4_critbeta} the intermediate results used in the
measurement of the relevant observables, as well as the resulting critical couplings.
We report both the measurements obtained with finite choices of  the plaquette subinterval, $\Delta_{u_p}$, along with the extrapolations, $\Delta_{u_p} \to 0$. The uncertainties quoted for each finite subinterval size are calculated by bootstrapping the results for the $n_R$ repeats. As explained elsewhere, we assume this estimate to capture in the stochastic error also the truncation error due to carrying out a finite number of RM iterations. The extrapolations are calculated by taking the results for each lattice volume and performing a fit linear in the square of the interval size $\Delta_{u_p}^2$. The errors quote for the extrapolations are the fitting errors.

\begin{table*}[t!]
\caption{ \label{tab:Nt4_critbeta} Measurements of the critical coupling in the  $Sp(4)$ pure gauge theory on lattices  with temporal size $N_t = 4$, obtained with  the LLR method. The plaquette subinterval  size is denoted as $\Delta_{u_p}$. Extrapolations to the limit of vanishing interval size are denoted by $\Delta_{u_p} \to 0$. The final extrapolations to the thermodynamic limit are denoted by $N_s \to \infty$.  
The coupling for which the free-energy on the meta-stable branches is equal is denoted as $\beta_{CV}(f)$, while $\beta_{CV}(C_V)$, $\beta_{CV}(B_V)$, and $\beta_{CV}(\chi_l)$, are the values of the couplings at the extrema of the specific heat, $C_V$, the Binder cumulant, $B_V$, and the Polyakov loop susceptibility, $\chi_l$, respectively. For thermodynamic limit  we include an additional bracket containing an estimation for the systematic error. 
}
\begin{center}
\begin{tabular}{|c|c|c|c|c|c|}
\hline
$N_s$ &  $\Delta_{u_p}$ & $\beta_{CV}(C_V)$ & $\beta_{CV}(B_V)$ & $\beta_{CV}(\chi_l)$ & $\beta_{CV}(f)$  \\
\hline
20 & 0.00064 & 7.340282(23) & 7.340246(24) & 7.339721(25) & 7.340102(25)\\
20 & 0.00048 & 7.340317(17) & 7.340283(17) & 7.339730(30) & 7.340136(19)\\
20 & $\to 0$ & 7.340362(49) & 7.340330(49) & 7.339742(75) & 7.340178(54)\\
24 & 0.00064 & 7.340111(19) & 7.340095(19) & 7.339800(20) & 7.340058(20)\\
24 & 0.00048 & 7.340113(14) & 7.340096(15) & 7.339788(18) & 7.340048(15)\\
24 & $\to 0$ & 7.340115(40) & 7.340097(41) & 7.339773(47) & 7.340036(43)\\
28 & 0.00048 & 7.340036(17) & 7.340028(16) & 7.339855(17) & 7.340055(16)\\
28 & 0.00025 & 7.340046(13) & 7.340035(12) & 7.339858(14) & 7.340064(14)\\
28 & $\to 0$ & 7.340050(19) & 7.340038(18) & 7.339859(21) & 7.340067(21)\\
40 & 0.00017 & 7.340011(9) & 7.340009(8) & 7.339978(8) & 7.340079(9)\\
40 & 0.00013 & 7.340011(9) & 7.340009(10) & 7.339976(10) & 7.340077(10)\\
40 & $\to 0$ & 7.340012(24) & 7.340009(24) & 7.339974(25) & 7.340074(24)\\
48 & 0.00013 & 7.340028(5) & 7.340026(5) & 7.340015(6) & 7.340084(5)\\
$\to \infty$&& 7.340018(7)(5) & 7.340019(7)(5) & 7.340052(8)(4) & 7.340089(8)(2)\\
\hline
\end{tabular}

\end{center}
\end{table*}

The extrapolation of the results in $\Delta_E^2$ has a theoretical basis, see Ref.~~\cite{Langfeld:2015fua}, and the scaling was demonstrated to be appropriate, for sufficiently small intervals, for the thermodynamic observables in the deconfinement transition in $SU(3)$ pure gauge theory, see Ref.~\cite{Lucini:2023irm}. We are however, unable to verify this assumption in this case as we only have two interval sizes points per lattice volume. The requirement of an additional limit, $\Delta_E^2 \to 0$, is a drawback of the algorithm, with the cost of additional interval sizes being weighed against additional lattice volumes and lattice spacings. The focus of this work was to pursue the thermodynamic limit and thus we opted to analyze two interval sizes per volume. Therefore, there is an additional systematic uncertainty, not quoted in the tables, due to taking the limit of vanishing interval size, which we estimate increases the total error by approximately $30\%$.

\section{Systematic uncertainty in the thermodynamic limit}
\label{App:Systematic}
To estimate the systematic uncertainty on the infinite volume extrapolations due to our choice of parametrization, we use the Akaike information criterion (AIC), Ref.~\cite{akaike1974new}. To each observable, we fit models with quadratic polynomials and linear polynomials, and exclude the smallest $N_{\text{cut}}$ data points. For each model we calculate the AIC term, given in Eq.~161 in Ref.~\cite{Borsanyi:2020mff}, and use this as a weight, $w_i \sim \exp(-\frac{1}{2}AIC)$. Further following the method presented in Sec.~21 of the supplementary information of Ref.~\cite{Borsanyi:2020mff}, we estimate the combined extrapolation and total error by constructing a cumulative distribution function given by $\sum w_i N(\mu_i, \sigma_i)$, where $N(\mu_i, \sigma_i)$ is a Gaussian distribution centered on the result of each extrapolation, $\mu_i$, with standard deviation given by the uncertainty on it, $\sigma_i$. From this we can get an estimate for the total error, including both stochastic and systematic sources, as the difference between the 16th and 84th percentile of the cumulative distribution, $\sigma_{tot} = \frac{1}{2}(y_{84} - y_{16})$. In the main body of the paper, Tabs.~\ref{tab:Nt4_critobs} and \ref{tab:Nt4_critbeta}, we present the results and error for our choice of model, $i^*$, based on the AIC and other physical motivations, and additional the increase in error due to the systematics as, $\mu_{i^*}(\sigma_{i^*})(\sigma_{tot} - \sigma_i)$.

\bibliography{bibliography}
\end{document}